\documentclass[12pt,preprint]{aastex}

\usepackage{graphics}
\usepackage{CJKutf8}

\shorttitle{Mapping NGC 1042}
\shortauthors{Rongxin et al.} 

\begin{document}

\title{The VIRUS-P Exploration of Nearby Galaxies (VENGA): Radial Gas Inflow and Shock Excitation in NGC 1042}

\author{
Rongxin Luo(\begin{CJK}{UTF8}{gbsn}罗荣欣\end{CJK})\altaffilmark{1,2}, 
Lei Hao\altaffilmark{1}, 
Guillermo A. Blanc\altaffilmark{3,4,5}, 
Shardha Jogee\altaffilmark{6}, 
Remco C. E. van den Bosch\altaffilmark{7}, 
Tim Weinzirl\altaffilmark{8}}

\altaffiltext{1}{Key Laboratory for Research in Galaxies and Cosmology, Shanghai Astronomical Observatory, 
Nandan Road 80, Shanghai, 200030, China, haol@shao.ac.cn}
\altaffiltext{2}{Graduate School of the Chinese Academy of Sciences, 19A, Yuquan Road, Beijing, China}
\altaffiltext{3}{Departamento de Astronom\'{i}a, Universidad de Chile, Camino del Observatorio 1515, Las 
Condes, Santiago, Chile}
\altaffiltext{4}{Centro de Astrof\'{i}sica y Tecnolog\'{i}as Afines (CATA), Camino del Observatorio 1515, 
Las Condes, Santiago, Chile}
\altaffiltext{5}{Visiting Astronomer, Observatories of the Carnegie Institution for Science, 
813 Santa Barbara Street, Pasadena, CA, 91101, USA}
\altaffiltext{6}{Department of Astronomy, University of Texas at Austin, 2515 Speedway, Stop C1400, 
Austin, TX 78712-1205, USA}
\altaffiltext{7}{Max-Planck-Institut f\"ur Astronomie, K\"{o}nigstuhl 17, D-69117 Heidelberg, Germany}
\altaffiltext{8}{School of Physics and Astronomy, The University of Nottingham, University Park, 
Nottingham, NG7 2RD, UK}

\begin{abstract}

NGC 1042 is a late type bulgeless disk galaxy which hosts a low 
luminosity Active Galactic Nuclei (AGN) coincident with a massive 
nuclear star cluster. In this paper, we present the 
integral-field-spectroscopy (IFS) studies of this galaxy, based 
on the data obtained with the Mitchell spectrograph on the 2.7 
meter Harlan J. Smith telescope. 
In the central $100\textrm{-}300\ \mathrm{pc}$ region 
of NGC 1042, we find a circumnuclear ring structure of gas with 
enhanced ionization, which we suggest is mainly induced by shocks. 
Combining with the harmonic decomposition analysis of the velocity 
field of the ionized gas, we propose that the shocked gas is the 
result of gas inflow driven by the inner spiral arms. The inflow 
velocity is $\sim 32\pm10\ \mathrm{km}\ \mathrm{s}^{-1}$ 
and the estimated mass inflow rate is 
$\sim 1.1\pm0.3 \times 10^{-3}\ \mathrm{M}_{\odot}\ \mathrm{yr}^{-1}$. 
The mass inflow rate is about one hundred times the blackhole's mass 
accretion rate ($\sim 1.4 \times 10^{-5}\ \mathrm{M}_{\odot}\ \mathrm{ yr}^{-1}$), 
and slightly larger than the star formation rate in the nuclear star 
cluster ($7.94 \times 10^{-4}\ \mathrm{M}_{\odot}\ \mathrm{yr}^{-1}$), 
implying that the inflow material is enough to feed both the AGN activity 
and the star formation in the nuclear star cluster. Our study highlights 
that secular evolution can be important in late-type unbarred galaxies 
like NGC 1042.

\end{abstract}


\section{Introduction}

\subsection{Feeding of Low Luminosity AGNs}

Since the discovery of the tight correlation between the masses
of super-massive black holes (SMBHs) and the global properties of 
their host galaxies \citep{Ferrarese2000,Gebhardt2000,Tremaine2002}, 
the co-evolution of galaxies and SMBHs has become an important 
topic in the study of the galaxy formation and evolution. One 
outstanding problem is to understand how the host galaxy can feed 
the central blackhole and trigger the nuclear activity. Theoretically, 
galaxies can affect the growth of blackholes via global interaction 
processes such as major or minor mergers, galaxy interactions and gas 
accretions \citep{Negroponte1983,Sanders1988,Barnes1991,Quinn1993,
Mihos1996,Kauffmann2000,DiMatteo2005,Hopkins2006,Hopkins2008}. 
The galaxy-wide secular processes driven by a series of internal 
non-axisymmetric structures (e.g. large-scale bars, spiral arms, 
ovals, nuclear spiral arms, and nested bars) can also play 
important roles in the growth of blackholes \citep[][and references 
therein]{Jogee2002,Jogee2005,Kormendy2004,Athanassoula2008,Sellwood2014}. 
These processes have been proposed to effectively dissipate the angular 
momentum and transfer material inward to the central regions of galaxies 
\citep[][and references therein]{Shlosman1989,Shlosman1990,Martini2004,Jogee2006}. 

Observationally, there have been many statistical 
studies to investigate the role of different feeding 
mechanisms. Studies of the links between the nuclear activity and 
the merger signatures (e.g., close pairs or disturbed hosts) have 
shown that, major mergers have limited role in triggering the nuclear 
activities \citep{Cisternas2011,Schawinski2010,Schawinski2011,Schawinski2012,
Weinzirl2011,Kocevski2012,Simmons2012,Simmons2013,Karouzos2014,Villforth2014} 
and may only dominate the feeding processes of the most luminous AGNs 
\citep{Kartaltepe2010,Koss2011,Treister2012,Rosario2012}. 
The role of secular processes has been mainly tested by exploring the
bar-AGN connection, which only presents marginal evidence for a direct
correlation \citep{Ho1997,Mulchaey1997,Hunt1999,Shlosman2000,Laine2002,
Laurikainen2004a,Oh2012,Lee2012,Cisternas2013,Cisternas2015,Cheung2015,
Galloway2015}. In addition, the dominant feeding mechanisms for AGNs of 
different luminosities and within different environments may be different 
\citep{Jogee2006,Kormendy2013}. 

Another way to explore the feeding mechanisms of AGNs is
to study the detailed properties of the interstellar medium (ISM)
in the vicinity of the SMBHs. The Hubble Space Telescope (HST) imaging 
have revealed various dust structures around AGNs, which is thought
to be a possible feeding channel of nuclear activity \citep{Malkan1998}. 
However, further studies about the matched samples of active and 
inactive galaxies do not find significant correlation between the
circumnuclear dust and the nuclear activities \citep{Martini2003}, 
although there is an exception for the early-type AGN 
hosts \citep{SimoesLopes2007}. 

With the advances of the integral field 
spectroscopy (IFS), the gas inflows have been directly observed 
from the two-dimensional gaseous velocity field for more and more 
nearby AGNs \citep{Fathi2006,Storchi-Bergmann2007,Riffel2008,
MuellerSanchez2009,Davies2009,vandeVen2010,SchnorrMueller2011,
Riffel2011,Riffel2013,Schnorr-Mueller2014,Schoenell2014}. 
These detailed studies of individual sources have shown 
that the inflowing material could be multi-phased, with typical 
inflow velocities of $50\textrm{-}100\ \mathrm{km}\ \mathrm{s}^{-1}$ 
and a large range in mass inflow rates 
($0.01\textrm{-}1\ \mathrm{M}_{\odot}\ \mathrm{yr}^{-1}$). While IFS 
observations mainly probe the ionized and warm molecular phase of 
the ISM, the gas inflows have also been detected in the cold molecular 
phase with the observations of millimeter/submillimeter interferometry 
\citep{Garcia-Burillo2012,Fathi2013,Combes2014}. Furthermore, the 
hints of kinematic differences have been observed in the local Seyfert 
galaxies. \citet{Dumas2007} found that the kinematics of the ionized gas 
in the circumnuclear regions of Seyfert galaxies show stronger perturbations 
than those in inactive galaxies. \citet{Hicks2013} reported that 
Seyfert galaxies have more concentrated stellar and H$_{2}$ surface 
brightness, lower stellar velocity dispersion, and elevated H$_{2}$ 
luminosity within $\sim 100\textrm{-}200\ \mathrm{pc}$ comparing to 
inactive galaxies. \citet{Davies2014} further showed that the circumnuclear 
disks of molecular gas widely exist in Seyfert galaxies. In these disks, 
the gas presents inflows and outflows kinematics superimposed on the disk 
rotations.

In this paper, we explore the feeding mechanism of the Low Luminosity 
AGN (LLAGN) NGC 1042, based on the IFS data taken from the VENGA survey 
(VIRUS-P Exploration of Nearby Galaxies) \citep{Blanc2013}. We are 
interested in this object because it is a late-type bulgeless galaxy 
with an accreting intermeidate-mass black hole \citep{Shields2008}. 
In recent years, LLAGNs have been found in the pesudobulges of late-type 
galaxies and even in the bulgeless/dwarf galaxies 
\citep[][and references therein]{Ho2008,Greene2012,Kormendy2013}. 
These blackholes generally have low masses ranging 
from $10{^4}\ \mathrm{M}_{\odot}$ to $10{^6}\ \mathrm{M}_{\odot}$ 
\citep{Greene2004,Greene2007,Dong2012,Reines2013}. The scaling relations 
between these blackholes and their host galaxies are less certain, and may 
show much larger scatters than those of SMBHs in the classical bulges and 
ellipticals\citep{Hu2008,Greene2010,Kormendy2011,Sani2011,McConnell2013,Kormendy2013}. 
This suggests their mass growth may not significantly correlate with the 
global processes in the hosts. Here, we use NGC 1042 as an example to study 
the mass growth of blackholes at the low mass end.

The paper is organized as follows. 
First, we describe the relevant properties of NGC 1042 in \S1.2. In \S2, 
we describe the observations and the data reduction. In \S3, we present 
the 2D maps of emission-line fluxes, emission-line ratios, the velocity 
field, and the velocity dispersions of ionized gas and stars in NGC 1042. 
We find a circumnuclear ring-like structure of ionized gas at 
$100\textrm{-}300\ \mathrm{pc}$ region that shows LINER-like emission. 
In \S4, we characterize the nature of this structure and quantify the 
kinematic properties of the ionized gas. We propose that this structure 
is the result of the gas moving inwards driven by the inner spiral arms 
of NGC 1042. In \S5, we calculate the mass inflow rate of the ionized 
gas and discuss its implications for the AGN feeding and star formation in 
the nuclear star cluster. We summarize in \S6.

\subsection{NGC 1042}

NGC 1042 is a nearby late-type \citep[morphological type SAB(rs)cd,][]{deVaucouleurs1991} 
bulgeless galaxy. The basic properties of this galaxy are listed in Table 
\ref{tbl-1}. The spiral arms in NGC 1042 are composed of two symmetric inner 
arms and multiple long and continuous outer arms, which are classified as 
$\mathrm{AC}\ 9$ in the classification of spiral arms in \citet{Elmegreen1987}. 
In the RC3 catalogue \citep{deVaucouleurs1991}, NGC 1042 is classified as a 
weakly barred (SAB) spiral galaxy. \citet{Marinova2007} performed ellipse 
fitting on the B and H band images of this galaxy and also classified it as 
barred galaxy (with a deprojected bar ellipticity of 0.6 and bar semi-major 
axis $4.4\ \mathrm{kpc}$ in the H-band). However, using a different approach 
of multi-component (bulge, bar, disk) decomposition, \citet{Weinzirl2009} 
found that NGC 1042 is best fit with a bulge and disk only, and re-classify 
it as unbarred. The unbarred classification is also proposed by some other 
works. \citet{Laurikainen2002} and \cite{Laurikainen2004} used Fourier 
techniques on the B,J,H and K band images and classified it as an unbarred 
galaxy. \citet{Buta2005} decomposed the perturbation strengths of gravitational 
potential induced by the bar and the spiral arms and proposed that the 
contribution of the spiral arms are dominant. They point out that the 
inner spiral arms sharply curve towards to the central region and produce 
a bar-like structure in NGC 1042, which can be misidentified as a bar. 
Based on the potential-density phase-shift method, \citet{Buta2009} confirmed 
their previous unbarred classification. In the Carnegie-Irvine Galaxy Survey, 
\citet{Li2011} also classify NGC 1042 as an unbarred galaxy based on the visual 
inspection, the geometric analysis and the Fourier method. 

The coincidence of a nuclear star cluster and an AGN has been found 
in NGC 1042. The nuclear star cluster was first identified by the 
isophotal analysis of the HST images \citep{Boeker2002}. After a 
detailed structural analysis, \citet{Boeker2004} found that the 
effective radius of the cluster is $\sim 0.02\arcsec$ ($1.94\ \mathrm{pc}$ 
at their adopted distance of $18.2\ \mathrm{Mpc}$). The dynamical 
mass and stellar population of the cluster were also presented 
by \citet{Walcher2005,Walcher2006} with the VLT UVES spectroscopy. 
The obtained dynamical mass of the cluster is 
$\sim 3 \times 10^6\ \mathrm{M}_{\odot}$. The observed mean cluster 
age and metallicity is $10^4\ \mathrm{Myr}$ and $Z = 0.02$, 
respectively. There is a young stellar population with stellar age 
of $\sim 10\ \mathrm{Myr}$ in the underlying old ($> 1\ \mathrm{Gyr}$) 
population of stars. The mass of this young stellar component is 
$\sim 7.94 \times 10^3\ \mathrm{M}_{\odot}$ \citep{Walcher2006}.

Based on the high-resolution optical spectrum done within the central
$1\arcsec \times 1\arcsec$ region, \citet{Shields2008} claimed that 
there exists a low-luminosity AGN (classified as LINER) coincident with 
the nuclear star cluster in NGC 1042. They found that the peak of the 
H$\alpha$ flux has an offset ($\sim 0.5\arcsec$) from the peak of the 
forbidden lines and the stellar continuum. They also found that the 
profile of the [NII]+H$\alpha$ emission lines is unusual: The [NII] 
lines deviate from Gaussian profiles and present high velocity wings 
extending to about $\pm 300\ \mathrm{km}\ \mathrm{s}^{-1}$ from the line 
center. Additionally, the full width at half maximum (FWHM) of the [NII] 
is observed to be larger than that of the H$\alpha$. They proposed that 
this unusual profile can be explained as the blended emission from a 
LLAGN and an offset HII region. The HII region dominates the H$\alpha$ 
emission and produces the offset of the peak of the H$\alpha$ flux. 
After removing the contamination from the adjacent HII region, they 
estimated the bolometric luminosity of the AGN to be 
$L_{bol} \sim 8 \times 10^{39}\ \mathrm{erg}\ \mathrm{s}^{-1}$ using the 
H$\alpha$ luminosity and use it to set a lower limit on the mass of the 
blackhole ($M_{BH} > 60\ \mathrm{M}_{\odot}$). \citet{Walcher2005,Walcher2006} 
found that the M/L of the nuclear star cluster from the dynamical 
estimation is smaller than that from the stellar population analysis. 
Considering that the M/L estimated from the stellar components do not 
include any contributions from the blackhole, \citet{Shields2008} concluded 
that the mass of the blackhole and the nuclear cluster should be of the 
similar order. They used $3 \times 10^6\ \mathrm{M}_{\odot}$ as the upper limit 
of the blackhole mass of NGC 1042. In addition, they used the velocity 
dispersion of the nuclear stellar cluster and the M-$\sigma$ relation 
to estimate the blackhole mass, which is $4 \times 10^5\ \mathrm{M}_{\odot}$, 
implying an intermediate-mass blackhole. 

Interestingly, the distance of NGC 1042 has dramatically different values 
in the literature, ranging from $4.21\ \mathrm{Mpc}$ \citep{Tully2008} 
to $18.2\ \mathrm{Mpc}$ \citep{Boeker2002}. There are mainly two independent 
methods to determine the distance of this galaxy. The first method 
is based on the heliocentric velocity and using the Hubble's law to 
calculate the redshift distance. The motion of the local universe 
is corrected in the calculation. This method depends on the selection 
of the $H_{0}$ value and the models of local motions. The redshift distance 
estimated by this method is found to be $16.7\ \mathrm{Mpc}$ by \citet{Tully1988}, 
$18.2\ \mathrm{Mpc}$ by \citet{Boeker2002}, and $13.2\ \mathrm{Mpc}$ by 
\citet{Theureau2007}. The former two estimations only correct for the infall 
of the Galaxy to the Virgo cluster, while \citet{Theureau2007} made a full 
model of the peculiar velocity to the Galaxy within $80\ \mathrm{Mpc}$. 
In addition, they also corrected the peculiar velocity of NGC 1042. The 
second method is to make use of the Tully-Fisher (TF) relation. This method 
is therefore independent with the redshift. Several earlier estimations 
using this method determined the distance of NGC 1042 at $\sim 8\ \mathrm{Mpc}$ 
($8.4\ \mathrm{Mpc}$ from \citealt{Tully1992}, $7.8\ \mathrm{Mpc}$, 
$7.77\ \mathrm{Mpc}$, $7.94\ \mathrm{Mpc}$ from J,H,K band estimations 
of \citealt{Theureau2007}), but a later estimation by \citet{Tully2008} 
places NGC 1042 at $4.21\ \mathrm{Mpc}$. This work adopts the Tully-Fisher 
relation calibrated in \citet{Tully2000} and shifts slightly its zero point 
to be consistent with the results of the HST Cepheid Key Project \citep{Freedman2001}. 
Interestingly, we find that the distance estimated by this method is very 
sensitive to the inclination angle adopted for the galaxy. \citet{Tully2008} 
used the B-band photometry to determine the inclination angle to be $\sim 57^{\circ}$. 
If we use the inclination angle of $37^{\circ}$ which is indicated by the 
2MASS photometry, and use the same method of \citet{Tully2008}, we find 
the distance of NGC 1042 is $\sim 8\ \mathrm{Mpc}$. If we use the kinematic 
inclination ($38.7^{\circ}$) measured from the harmonic decomposition modelling 
of this paper (see Appendix B of this paper), the obtained distance changes 
to $\sim 7.3\ \mathrm{Mpc}$. Since the large scatter of the distance estimations 
of NGC 1042 may require further exploration, which is beyond the focus of this 
paper, we choose the estimation from the latest literature \citep{Tully2008}, 
adopting $4.21\ \mathrm{Mpc}$ as the distance of NGC 1042 in this paper. It is 
also the smallest distance estimation of all in literature so far, so our analyses 
can be considered as a reference for others by simply scaling up. We remind the 
readers to be cautious on the big uncertainties of the distance estimation of NGC 1042.

\section{Observations and Data Reduction}

\subsection{Observations}

Observations of NGC 1042 were carried out as part of the VENGA 
survey \citep{Blanc2013} which is a wide-field optical 
($3600\textrm{-}6800\ \mathrm{\AA}$) IFS survey of 30 nearby spiral 
galaxies. It uses the Mitchell spectrograph 
\citep[formerly known as VIRUS-P,][]{Hill2008a} on the 2.7 meter 
Harlan J. Smith telescope to observe about 44000 individual spectra 
across the disks of these objects. As a fiber-fed integral field 
unit (IFU), Mitchell consists of 246 fibers sampling 
a $1.7\arcmin \times 1.7\arcmin$ field of view with a 1/3 filling factor. The 
diameter of each fiber is $200\ \mu m$ corresponding to $4.24\arcsec$ on 
the sky. The VP1 grating was used in a blue and a red setup to cover 
from $3600\ \mathrm{\AA}$ to $6800\ \mathrm{\AA}$ spectral range with 
a spectral resolution of $5\ \mathrm{\AA}$ (FWHM). 

\begin{figure}[htbp]
\begin{center}
\includegraphics[width=0.55\textwidth]{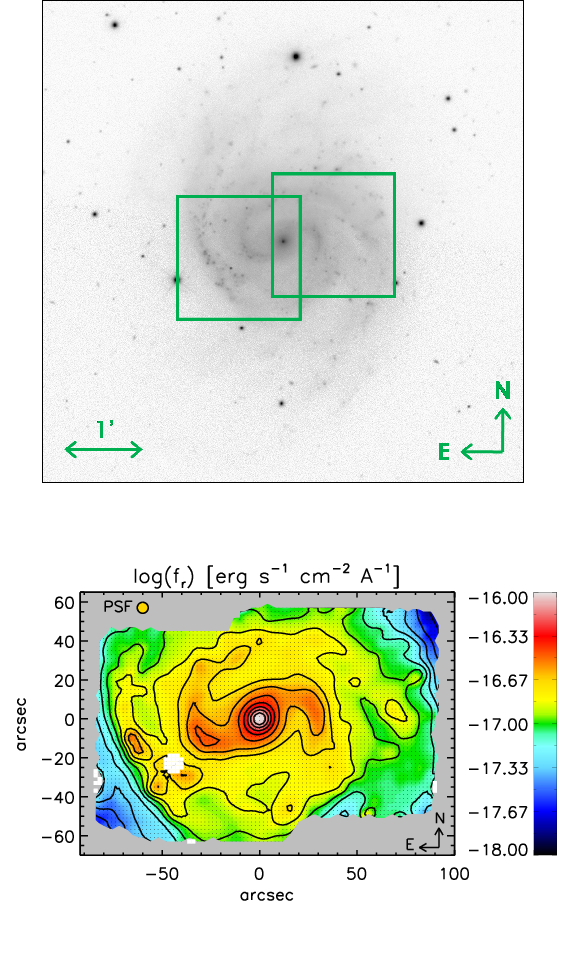}
\caption{Top panel: The SDSS r-band image of NGC 1042. The 
green squares present two pointings of the Mitchell observations.
Bottom panel: The reconstructed r-band flux map of NGC 1042.
Black dots mark the position of each spaxel in the datacube.
Black contours show isophotes of the SDSS r-band image. The 
foreground stars are masked as blank regions (see \S2.2). 
The gold circle in the top left corner presents the 
$5.6\arcsec$ FWHM PSF of the VENGA data.}
\label{fig-ngc1042rvp}
\end{center}
\end{figure}

Figure \ref{fig-ngc1042rvp} shows the r-band SDSS image of NGC 1042, 
overlaid with the area covered by the Mitchell observations. Two 
slightly offset pointings of the Mitchell are used to map this 
galaxy across a $3.5\ \mathrm{kpc} \times 2.5\ \mathrm{kpc}$ area. 
For each pointing, three dithers are observed at relative positions
$(\Delta \alpha, \Delta \delta) = (0.0'',0.0'')$, $(-3.6'',-2.0'')$, 
and $(0.0'',-4.0'')$ from the origin, to fill the whole area of 
the Mitchell field of view. The observation of each dither is divided 
into several science frames with exposure times ranging from 20 min to 
30 min. The 5 min off-source frames are observed between each science 
frame, which are used to perform sky subtraction. The calibration frames 
(bias, arc lamps, twilight flats, and spectrophotometric standard stars) 
are also observed in the same night. A detailed description of the VENGA 
observing strategy is presented in \citet{Blanc2013}.

The observational details of NGC 1042 are summarized in Table 
\ref{tbl-2}. For each dither in each pointing we list the total 
on-source exposure time, the number of science frames, the average
seeing of the frames, and the mean atmospheric transparency. NGC 1042 
was observed in five runs of the VENGA survey. Observing conditions 
were variable between different runs and within different nights 
during the same observing run, ranging from photometric to partly 
cloudy with average atmospheric transparency down to $\sim 60\%$. 
17 out of 76 frames ($22\%$) are rejected due to the pointing 
errors or bad sky subtraction problems. The seeing of the remaining 
good frames ranged from $1.63\arcsec$ to 
$2.93\arcsec$, with a median at $2.20\arcsec$. 

Overall we spent 27 hours of exposure time on the 59 good frames of 
this galaxy. As a result, the spectra have high S/N ratios per spectral 
resolution element, with a median value at $\sim 100$ in the continuum. 
In the central parts ($500\ \mathrm{pc} \times 500\ \mathrm{pc}$) of the 
galaxy we typically have $\mathrm{S/N} > 200$, while the spectra in the 
most outer regions (from $2.0\ \mathrm{kpc}$ to $2.5\ \mathrm{kpc}$) have 
a median $\mathrm{S/N} \sim 55$. We only have 20 out of 4789 
spaxels ($0.42\%$) with S/N less than 10. They mainly locate at the right 
top corner of the galaxy as shown in Figure \ref{fig-ngc1042rvp}.

\subsection{Data Reduction}

The details of the data processing and spectral analysis of the VENGA 
survey have been described in \citet{Blanc2013}. As part of this 
survey, we reduce and analyse the data of NGC 1042 in the same way. 

All raw data are reduced by the VACCINE pipeline \citep{Adams2011,
Blanc2013}. VACCINE performs bias and overscan subtraction, flat fielding, 
cosmic-ray rejection, wavelength calibration, and sky subtraction. Then 
the frames of spectrophotometric standard stars are used to perform 
the relative flux calibration for the 2D spectra in each observing run. 
Afterwards, the broad-band images reconstructed from these 2D spectra 
are compared with the SDSS broad-band images to perform the absolute 
flux calibration and astrometry. The flux errors are obtained by combining 
the read-noises and Poisson uncertainties in the CCD processing, and then 
propagated throughout the rest of the data reduction. The instrumental
spectral resolution is measured from the emission lines in the arc lamp 
frames and fitted as a function of wavelength. All these information are 
combined together to generate a datacube and a row-stacked spectra (RSS) 
file of NGC 1042. 

In these datasets, the spectra are re-sampled to a common 
wavelength grid at $1.1\ \mathrm{\AA}$ for regularly linear sampling and 
$(\Delta\lambda/\lambda)\ c = 60\ \mathrm{km}\ \mathrm{s}^{-1}$ for regularly 
logarithmic sampling. The spaxel has a regular scale of $2\arcsec \times 2\arcsec$, 
which can roughly Nyquist sample the final spatial PSF. For each spaxel the final 
spectrum is obtained by combining the spectra of surrounding fibers using a 
Gaussian spatial filter and adopting an inverse variance weighting scheme. 
The FWHM of the Gaussian filter is $4.24\arcsec$, matching the fiber size. 
The final spatial PSF is obtained by convolving the top-hat profile 
of the fiber ($4.24\arcsec$ diameter) with the seeing ($\sim 2.20\arcsec$) and 
the Gaussian filter ($4.24\arcsec$ FWHM). This PSF can be well described as a 
Gaussian with $5.6\arcsec$ FWHM. At the distance of $4.21\ \mathrm{Mpc}$ for 
NGC 1042, it corresponds to $112\ \mathrm{pc}$. For a detailed description 
of the data combining process and data format, see \S4.7 of \citet{Blanc2013}. 

The spectral analysis are performed by the PARADA pipeline, which involves the 
pPXF \citep{Cappellari2004} and GANDALF \citep{Sarzi2006} softwares as the main 
fitting procedures. Before fitting the spectra, the Galactic extinctions are
corrected by adopting a Milky Way extinction law as parametrized by \citet{Pei1992} 
and the extinction values from the maps of \citet{Schlegel1998}. The pPXF 
is used to fit the stellar continuum. The stellar line-of-sight velocity 
($\upsilon\ast$) and the velocity dispersion ($\sigma\ast$) are estimated 
from the fit. The spectrum of each spaxel is fitted individually with a 
linear combination of stellar templates convolved with a line-of-sight velocity 
dispersion (LOSVD) by using the “penalized pixel” technique \citep{Cappellari2004}. 
The stellar templates are the same as those in \citet{Blanc2013}. These templates 
are selected from MILES stellar library version 
9.1 \citep{Sanchez-Blazquez2006,Falcon-Barroso2011} 
and include 48 stars which span a wide range in spectral types (O through M), 
luminosity classes (I through V), and metallicities ($-2 < \mathrm{[Fe/H]} < 1.5$).
The spectral resolution of the templates and the observed spectra are matched 
to the worst instrumental resolution at any given wavelength in the datacube, 
which translates in a final spectral resolution of 
$110\ \mathrm{km}\ \mathrm{s}^{-1}$ in the red setup and 
$200\ \mathrm{km}\ \mathrm{s}^{-1}$ in the blue setup. The LOSVD can be 
parametrized by a Gauss-Hermite polynomial in pPXF, but we only fit for 
the first two moments $\upsilon\ast$ and $\sigma\ast$, which means that 
we actually fit a Gaussian LOSVD. 

After constraining the stellar kinematics, we measure the emission-line properties 
with the GANDALF. GANDALF fits the full spectrum by re-computing the weights 
given to the different stellar templates and adding Gaussian profiles to model the 
emission lines. The stellar line-of-sight velocity ($\upsilon\ast$) and the velocity 
dispersion ($\sigma\ast$) derived in the pPXF fitting are fixed in the GANDALF fitting. 
We tie the kinematics of all emission lines to a common set of parameters 
($\upsilon_{gas}$, $\sigma_{gas}$) in order to improve the fitting quality of faint 
lines. This ensures that the kinematic parameters obtained are mostly constrained by 
the brightest emission lines in the spectrum (typically H$\alpha$, [OIII]$\lambda$5007, 
and [OII]$\lambda$3727). The errors of emission line parameters are estimated based on 
the covariance matrix computed during the non-linear Levenberg-Marquardt fitting. 
In Table \ref{tbl-3}, we report the median S/N of each emission line over 
all spaxels in the datacube and the fraction of the observed area in which the 
emission lines are detected at $5\ \sigma$ and $3\ \sigma$. We detect H$\alpha$ at 
$5\ \sigma$ over $98\%$ of the full datacube with a median $\mathrm{S/N} = 38.9$. 
Most of the strong lines ([NII]$\lambda\lambda$6548,6583, [SII]$\lambda\lambda$6717,6731, 
H$\beta$, [OIII]$\lambda$5007, [OII]$\lambda\lambda$3726,3729 doublets) are detected 
at $3\ \sigma$ over more than $\sim 80\%$ of the observed area. 

We construct a series of 2D maps of NGC 1042, including the emission-line flux 
maps, the emission-line ratio maps, the maps of the velocity field, and the maps of 
the velocity dispersions. We identify the foreground Milk Way stars from the visual 
inspection of the stellar velocity field and the SDSS images. In all maps, we flag 
the spaxels contaminated by these stars and mask the surrounding spaxels within 
$5\arcsec$ as blank regions. For the emission-line flux and ratio maps 
(Figure \ref{fig-flux} and Figure \ref{fig-lineratio}), we adopt a S/N cut of 
5 for each line used in the corresponding maps to select the regions with reliable 
measurements. The spaxels that do not satisfy the above S/N cut are masked with 
blank regions. In the maps of the velocity field and the velocity dispersion of 
the ionized gas (Figure \ref{fig-vgas}), the regions having a velocity error or 
a velocity dispersion error larger than $10\ \mathrm{km}\ \mathrm{s}^{-1}$ are 
masked as blank regions.   

\section{Results} 

\subsection{Maps of the Emission Line Fluxes and the Line Ratios}

\begin{figure}[htbp]
\begin{center}
\includegraphics[width=\textwidth]{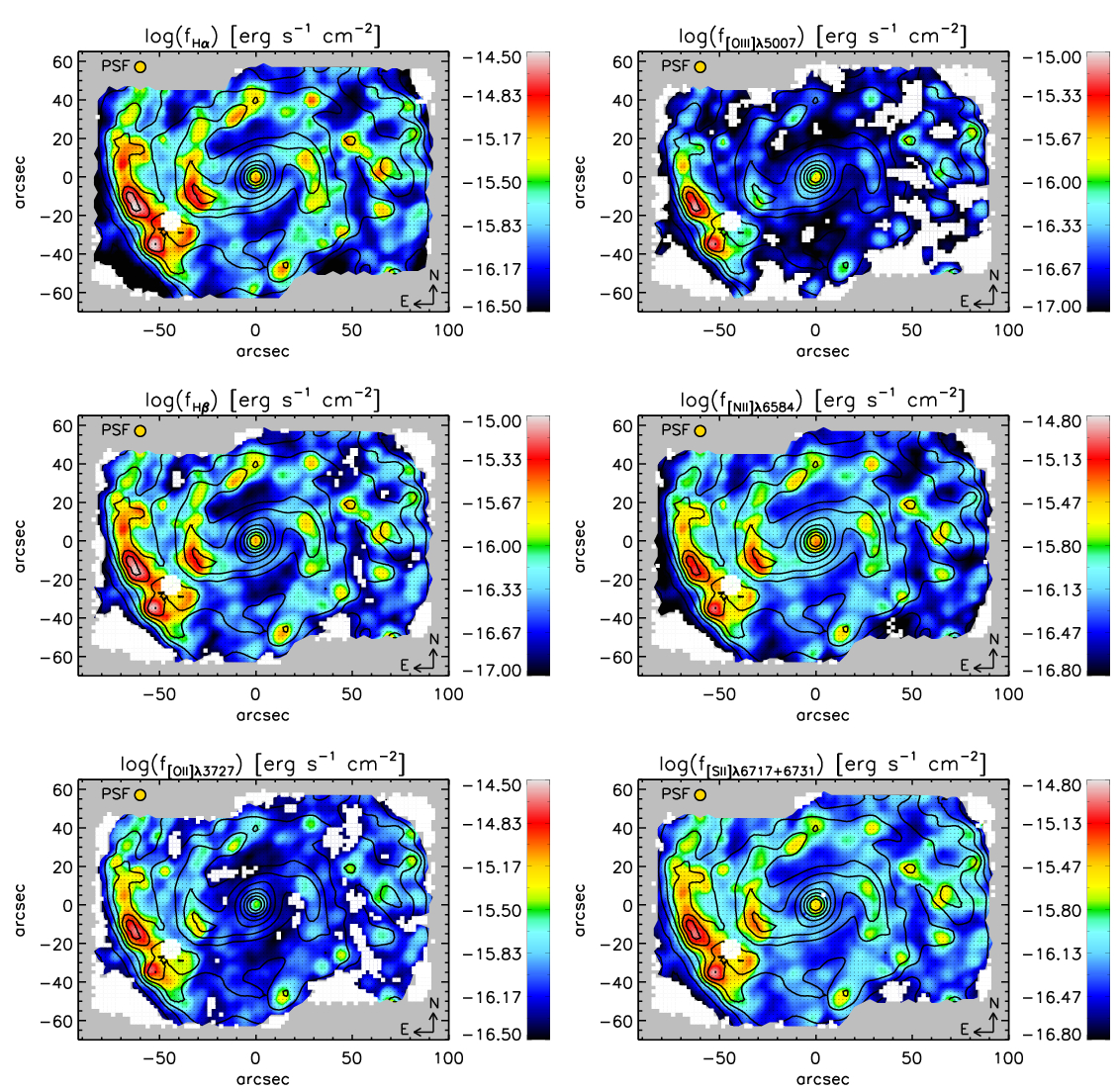}
\caption{The emission line flux maps of H$\alpha$, H$\beta$, 
[OII]$\lambda$3726+$\lambda$3729, [OIII]$\lambda$5007, 
[NII]$\lambda$6584 and [SII]$\lambda$6716+$\lambda$6731 in NGC 1042. 
The spaxels contaminated by the foreground stars (see \S2.2) and having 
line S/N below 5 are masked as blank regions. The gold circles 
in the top left corners present the $5.6\arcsec$ FWHM PSF of the VENGA 
data. The black contours represent the SDSS r-band isophotes.}
\label{fig-flux}
\end{center}
\end{figure}

\begin{figure}[htbp]
\begin{center}
\includegraphics[width=\textwidth]{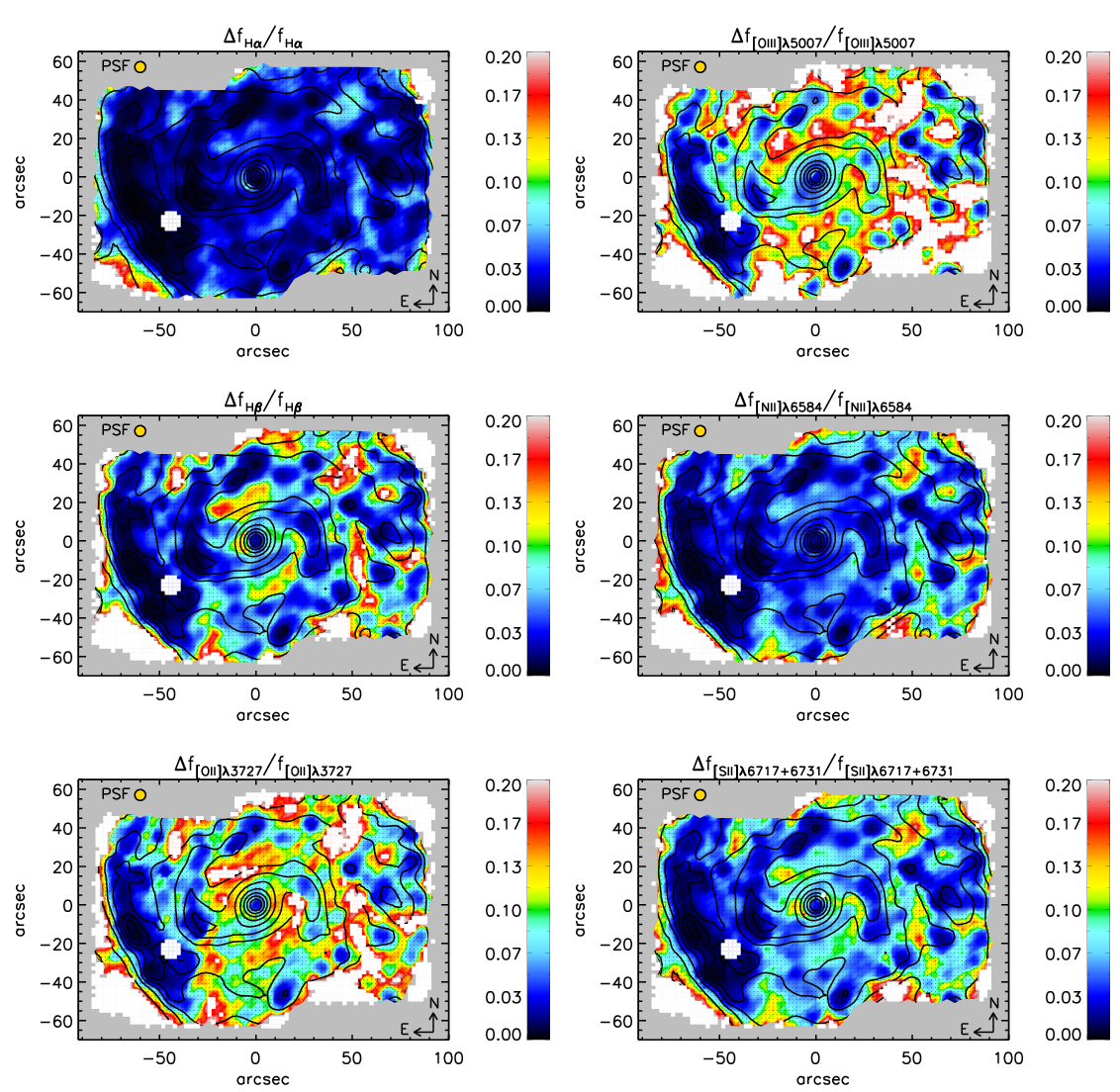}
\caption{The error maps of the emission line fluxes from 
H$\alpha$, H$\beta$, [OII]$\lambda$3726+$\lambda$3729, 
[OIII]$\lambda$5007, [NII]$\lambda$6584 and 
[SII]$\lambda$6716+$\lambda$6731 in NGC 1042. The spaxels 
contaminated by the foreground stars (see \S2.2) and having 
line S/N below 5 are masked as blank regions. The gold 
circles in the top left corners present the $5.6\arcsec$ 
FWHM PSF of the VENGA data. The black contours represent 
the SDSS r-band isophotes.}
\label{fig-fluxerror}
\end{center}
\end{figure}

Figure \ref{fig-flux} presents the maps of emission-line fluxes of 
H$\alpha$, H$\beta$, [OII]$\lambda$3726+$\lambda$3729, [OIII]$\lambda$5007, 
[NII]$\lambda$6584 and [SII]$\lambda$6716+$\lambda$6731 in NGC 1042. 
Figure \ref{fig-fluxerror} illustrates the errors corresponding to 
these flux measurements. The morphology of the Balmer-linestrengths
and the forbidden-linestrengths agree very well, all show clear spiral arms. 
The overall distribution of these emissions is significantly asymmetric. 
The emission in the eastern arms is stronger than that in the western 
arms, suggesting that the eastern arms contain more active star-forming 
regions. In the inter-arm regions we can detect significant amounts of 
emission, with a surface brightness that is one to two orders of magnitude 
fainter than that in the arms. There is also an enhanced emission in the center.

\begin{figure}[htbp]
\begin{center}
\includegraphics[width=\textwidth]{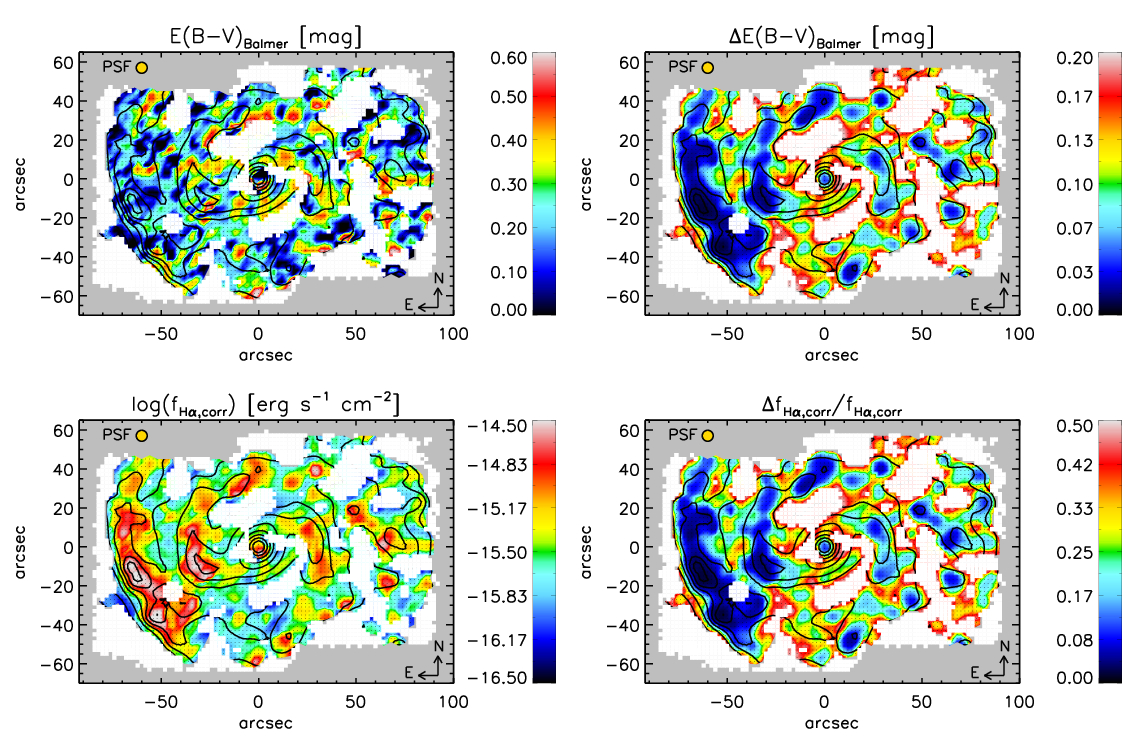}
\caption{Top panel: The map of dust reddening $E(B-V)$ and 
its error in NGC 1042. We masked the spaxels with error of 
$E(B-V)$ larger than $0.2\ \mathrm{mag}$ as blank regions, 
which roughly amounts to an uncertainty of $\sim 50\%$ in 
the extinction correction factor at the wavelength of H$\alpha$. 
Bottom panel: The map of extinction corrected H$\alpha$ fluxes 
and their relative errors in NGC 1042. In both maps, the spaxels 
contaminated by the foreground stars are also masked (see \S2.2). 
The gold circles in the top left corners present the $5.6\arcsec$ 
FWHM PSF of the VENGA data. The black contours represent the SDSS 
r-band isophotes.} 
\label{fig-extinction}
\end{center}
\end{figure}

The amount of dust reddening can be determined by comparing 
the observed H$\alpha$/H$\beta$ ratio with the intrinsic 
H$\alpha$/H$\beta$ ratio expected from the recombination 
theory. We assume an intrinsic H$\alpha$/H$\beta$ ratio of 2.87 
\citep{Osterbrock2006} and a Milk Way extinction law 
parameterized by \citet{Pei1992}. Figure \ref{fig-extinction} presents 
the map of $E(B-V)$ and the extinction-corrected H$\alpha$ flux. The 
measured errors of $E(B-V)$ are obtained by propagating the 
uncertainties in the H$\alpha$ and H$\beta$ line fluxes. In Figure 
\ref{fig-extinction} we masked the spaxels with error of $E(B-V)$ larger 
than $0.2\ \mathrm{mag}$, which roughly amounts to an uncertainty of 
$\sim 50\%$ in the extinction correction factor at the wavelength of H$\alpha$.
The median extinction of NGC 1042 is about $A_{V} = 0.72\ \mathrm{mag}$ 
(assuming $R_{V} = A_{V}/E(B-V) = 3.1$). Considering many spaxels have 
large errors in $E(B-V)$ because of weak H$\beta$ fluxes, we decide 
not to perform the extinction corrections for the analysis in the 
remaining of the paper since we mainly focus on the emission-line ratios 
from now on. The effect of dust reddening will not significantly influence 
the values of emission line ratios we are interested in, as the central 
wavelengths of these lines are relatively close.

\begin{figure}[htbp]
\begin{center}
\includegraphics[width=\textwidth]{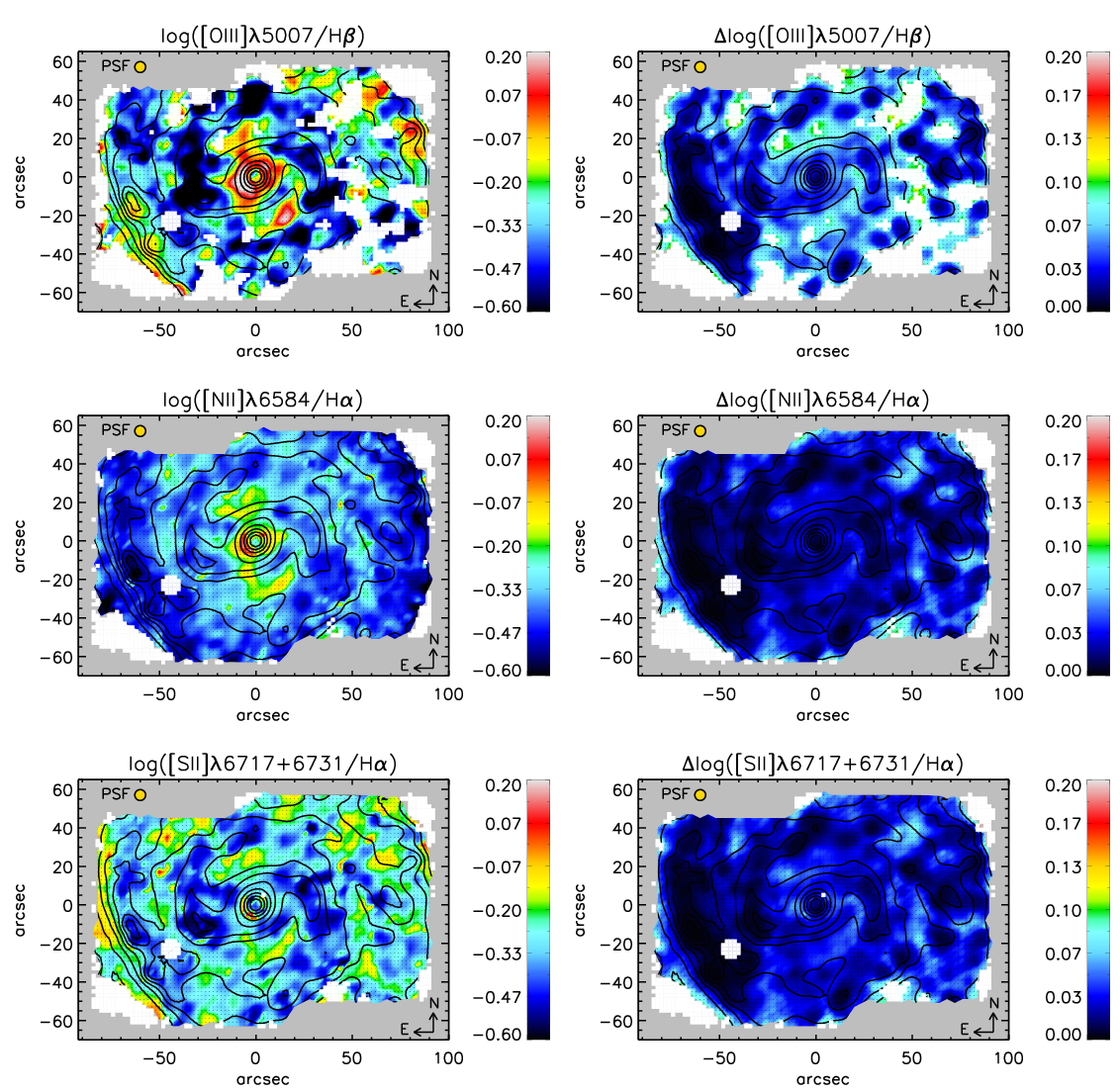}
\caption{The error maps of the emission line ratios, 
[OIII]$\lambda$5007/H$\beta$, [NII]$\lambda$6584/H$\alpha$, 
and [SII]$\lambda$6716+$\lambda$6731/H$\alpha$ in NGC 1042. 
In each map, the spaxels contaminated by the foreground stars 
are masked as blank regions (see \S2.2). The spaxels are also 
masked as blank regions if any emission line used in each map 
have $\mathrm{S/N} < 5$. The gold circles in the top 
left corners present the $5.6\arcsec$ FWHM PSF of the VENGA 
data. The black contours represent the SDSS r-band isophotes.}
\label{fig-lineratio}
\end{center}
\end{figure}

Figure \ref{fig-lineratio} presents the maps of 
[OIII]$\lambda$5007/H$\beta$, [NII]$\lambda$6584/H$\alpha$, 
and [SII]$\lambda$6716+$\lambda$6731/H$\alpha$. The line ratios 
of [OIII]$\lambda$5007/H$\beta$ and [NII]$\lambda$6584/H$\alpha$ 
are sensitive to the ionization state and metallicity of the ionized 
gas. The line ratio of [SII]$\lambda$6716+$\lambda$6731/H$\alpha$
traces a combination of temperature, ionization state and metallicity 
of ionized gas \citep{Blanc2015}. The higher values of these line 
ratios means a higher ionization state of the gas. 
The higher [OIII]$\lambda$5007/H$\beta$,
[NII]$\lambda$6584/H$\alpha$ and [SII]$\lambda$6716+$\lambda$6731/H$\alpha$
are detected in the inter-arm regions, with $\sim 0.3\ \mathrm{dex}$ greater 
than those detected in the arms. This can be attributed to the existence 
of the diffuse ionized gas (DIG). The DIG is a layer of nearly fully 
ionized hydrogen around bright HII regions, which can extend to the 
inter-arm regions and several kpc above and below the mid-plane of 
disk galaxies \citep{Mathis2000,Haffner2009,Blanc2009}. It has been 
considered as a major component of the interstellar medium of the 
Milky Way and other disk galaxies (see reviews by \citealt{Mathis2000} 
and \citealt{Haffner2009}). The main ionization source of the DIG is the 
Lyman continuum photons leaking from the HII regions associated with the 
hot stars \citep[][and references therein]{Haffner2009}. In DIG the 
detected Balmer and forbidden emission are more diffused and the observed 
line ratios of [OIII]$\lambda$5007/H$\beta$, [NII]$\lambda$6584/H$\alpha$ 
and [SII]$\lambda$6716/H$\alpha$ are always enhanced 
\citep[Kaplan et al. submitted]{Bland-Hawthorn1991,Haffner1999,Reynolds1999,
Hoopes2003,Madsen2006,Voges2006,Blanc2009}.

\begin{figure}[htbp]
\begin{center}
\includegraphics[width=0.9\textwidth]{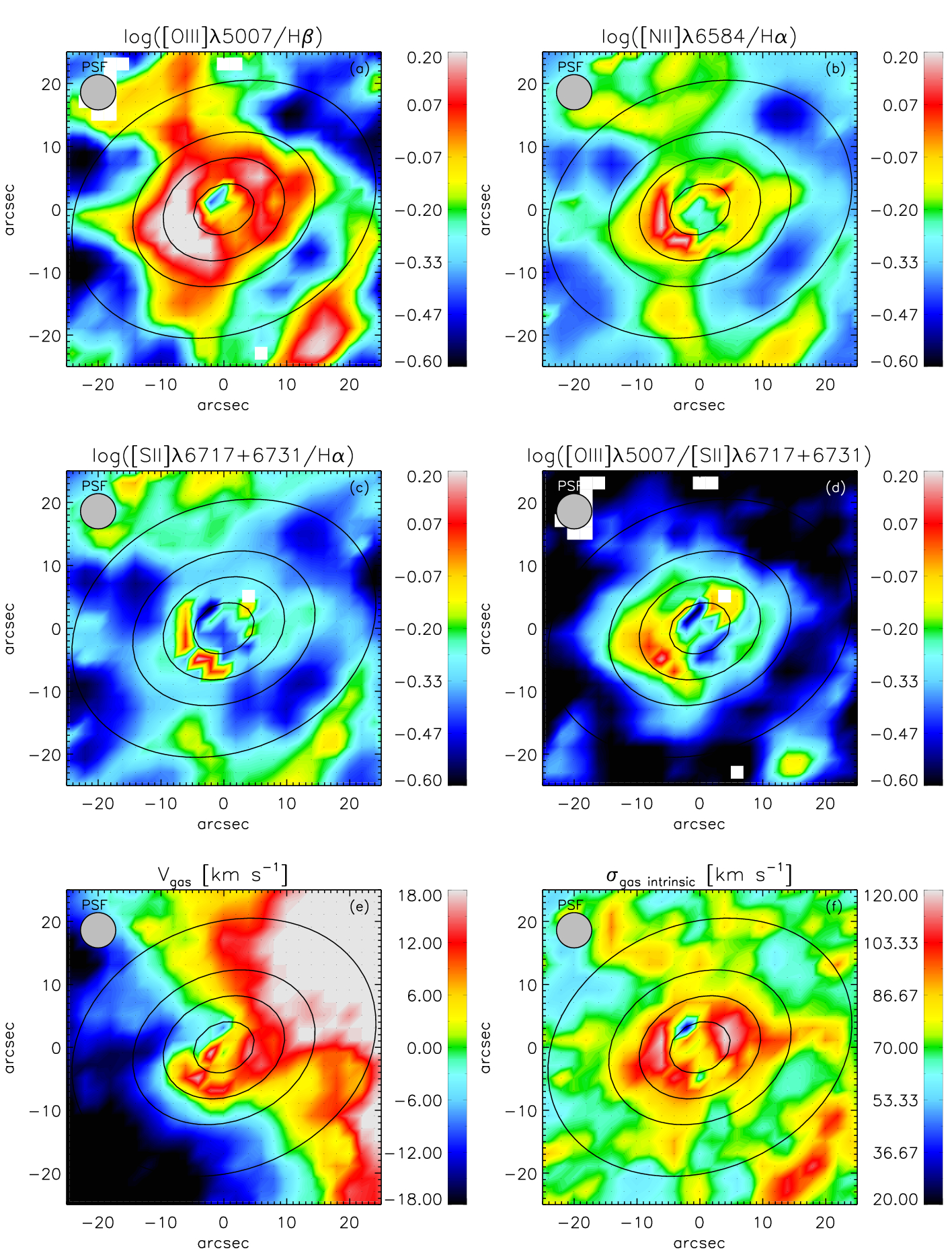}
\caption{The zoomed-in maps of the emission-line ratios, the velocity 
field and the velocity dispersion of the ionized gas in the central 
$500\ \mathrm{pc} \times 500\ \mathrm{pc}$ of NGC 1042. The spaxels 
are masked as blank regions if any emission line used in each map have 
$\mathrm{S/N} < 5$. The gray circles in the top left corners present 
the $5.6''$ FWHM PSF of the VENGA data. The black contours represent 
the constant radii of 0.1, 0.2, 0.3 and $0.5\ \mathrm{kpc}$.} 
\label{fig-lineratiozoom}
\end{center}
\end{figure}

In addition to the DIG component, we also find a Circumnuclear 
Ring-like Ionized Gas Structure (hereafter, we use an abbreviate name 
as CRIGS) that shows enhanced line ratios in the central region. We 
show the zoomed-in maps of the emission line ratios in the central 
$500\ \mathrm{pc} \times 500\ \mathrm{pc}$ region in Figure \ref{fig-lineratiozoom}. 
This structure extends from $100\ \mathrm{pc}$ to $300\ \mathrm{pc}$ from the center. 
The values of [OIII]$\lambda$5007/H$\beta$, [NII]$\lambda$6584/H$\alpha$ and 
[SII]$\lambda$6716+$\lambda$6731/H$\alpha$ in the CRIGS are $\sim 0.5\ \mathrm{dex}$ 
higher than the typical values seen in the spiral arm, and $\sim 0.2\ \mathrm{dex}$ 
higher than the typical values of DIG in the inter-arm regions. The enhanced 
[OIII]$\lambda$5007/H$\beta$ in this location has also been observed by the 
SAURON survey \citep{Ganda2006}. This ring structure is different from the 
well-known circumnuclear star-forming rings in some galaxies that are located 
at several hundreds pc from the center \citep{Comeron2010,Comeron2014}. 
These star-forming rings show enhanced star formation, thus, enhanced Balmer 
emission-line fluxes compared to their disk regions. For the 
CRIGS in NGC 1042, neither H$\alpha$ nor H$\beta$ fluxes are enhanced. On 
the contrary, they show slight deficits. The [OIII]$\lambda$5007, 
[NII]$\lambda$6584, and [SII]$\lambda$6716+$\lambda$6731 fluxes are also
slightly weaker in the CRIGS, but the line ratios of [OIII]$\lambda$5007/H$\beta$, 
[NII]$\lambda$6584/H$\alpha$ and [SII]$\lambda$6716+$\lambda$6731/H$\alpha$ are 
enhanced compared to other regions of NGC 1042.

\subsection{Maps of the Velocity Field and the Velocity Dispersion of Ionized Gas}

\begin{figure}[htbp]
\begin{center}
\includegraphics[width=\textwidth]{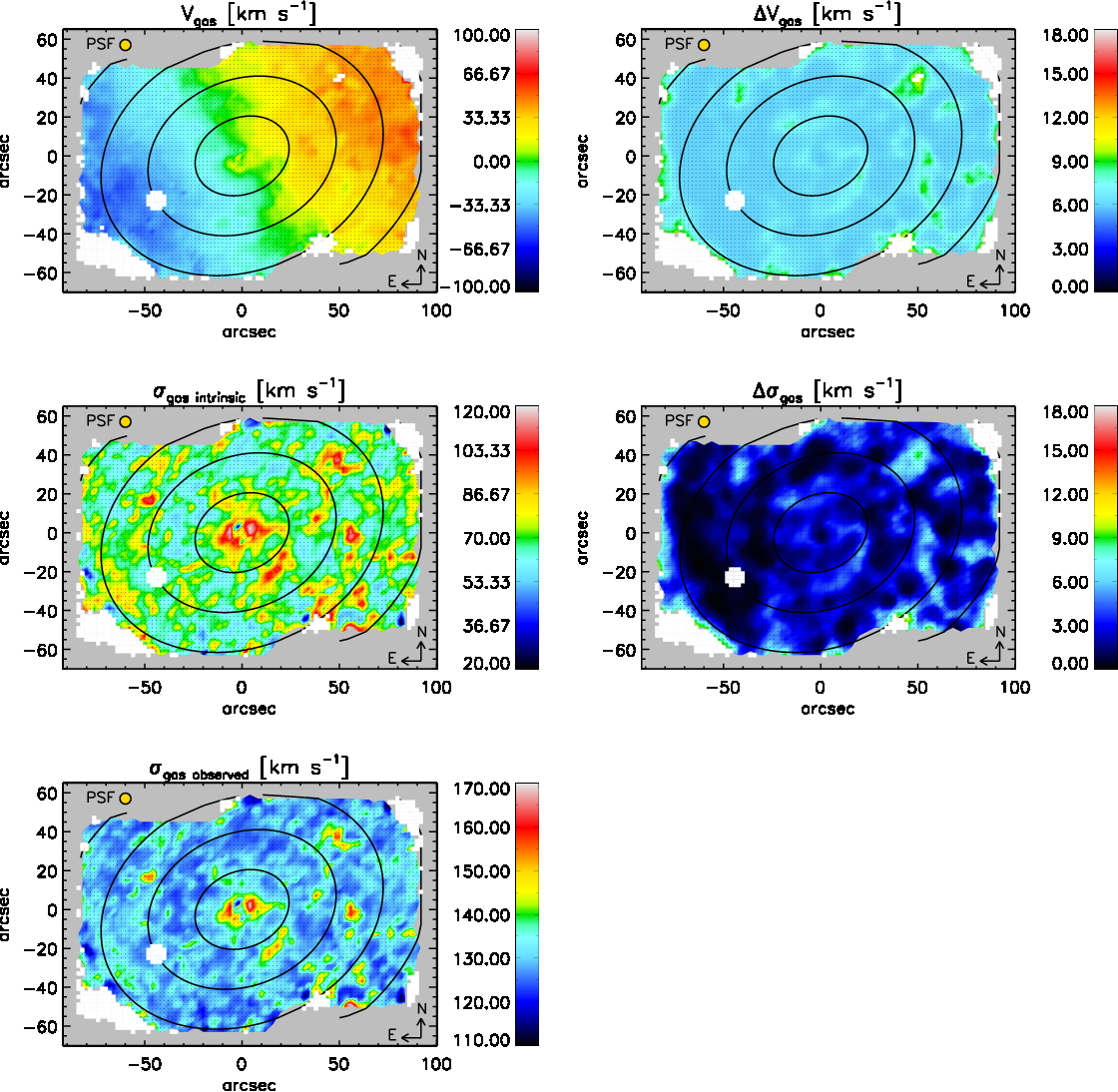}
\caption{The maps of the velocity field and the velocity dispersion (both 
observed and intrinsic) of the ionized gas in NGC 1042 and their corresponding 
error maps. The spaxels contaminated by the foreground stars and having velocity 
error (or velocity dispersion error) larger than $10\ \mathrm{km}\ \mathrm{s}^{-1}$ 
are masked. The gold circles in the top left corners present the $5.6\arcsec$ 
FWHM PSF of the VENGA data. The black contours represent the constant radii 
in steps of $0.5\ \mathrm{kpc}$.} 
\label{fig-vgas}
\end{center}
\end{figure}

In Figure \ref{fig-vgas} we present the maps of the velocity field and 
the velocity dispersion of the ionized gas in NGC 1042. The error associated 
with the velocity is a combination of the errors in the emission line 
measurements and in wavelength calibrations. The errors for the velocity 
dispersion are obtained from the GANDALF fitting.

In addition to the map of the observed velocity dispersion, we also show the 
map of the intrinsic velocity dispersion of the ionized gas, which is obtained 
by subtracting the instrumental spectral resolution of the red setup 
($\sim 110\ \mathrm{km}\ \mathrm{s}^{-1}$, see section \S2.2) in quadrature 
from the observed velocity dispersion. \citet{Blanc2013} showed that the stellar 
velocity dispersion of the VENGA galaxies can be largely overestimated when the 
intrinsic velocity dispersion is significantly smaller than the instrumental 
resolution, using the Monte Carlo simulations. We do not perform similar analysis 
to the velocity dispersion of the ionized gas, and it could also be overestimated 
for the regions with relatively low intrinsic velocity dispersion. However, for the 
regions with intrinsic velocity dispersion comparable to or larger than the 
instrumental resolution, our measurements are robust.

In general, the ionized gas follows the regular circular rotation. Some extent 
of twists exists in the central $500\ \mathrm{pc} \times 500\ \mathrm{pc}$ region. 
The velocity dispersion in the inter-arm region is relatively higher than that 
in the spiral arm, while some bright HII regions also show enhanced velocity 
dispersions. As shown in Figure \ref{fig-lineratiozoom} (e) and (f), the ionized 
gas in the central $500\ \mathrm{pc} \times 500\ \mathrm{pc}$ region presents a 
strongly spiral-like kinematic twist and a significant enhancement of velocity 
dispersion in the CRIGS, reaching to 90-130 $\mathrm{km}\ \mathrm{s}^{-1}$. 
This is higher than the typical values of the ionized gas in spiral galaxies 
\citep[$35\ \mathrm{km}\ \mathrm{s}^{-1}$,][]{Epinat2008}.

\subsection{The LINER-like emission in the CRIGS of NGC 1042}

\begin{figure}[htbp]
\begin{center}
\includegraphics[width=\textwidth]{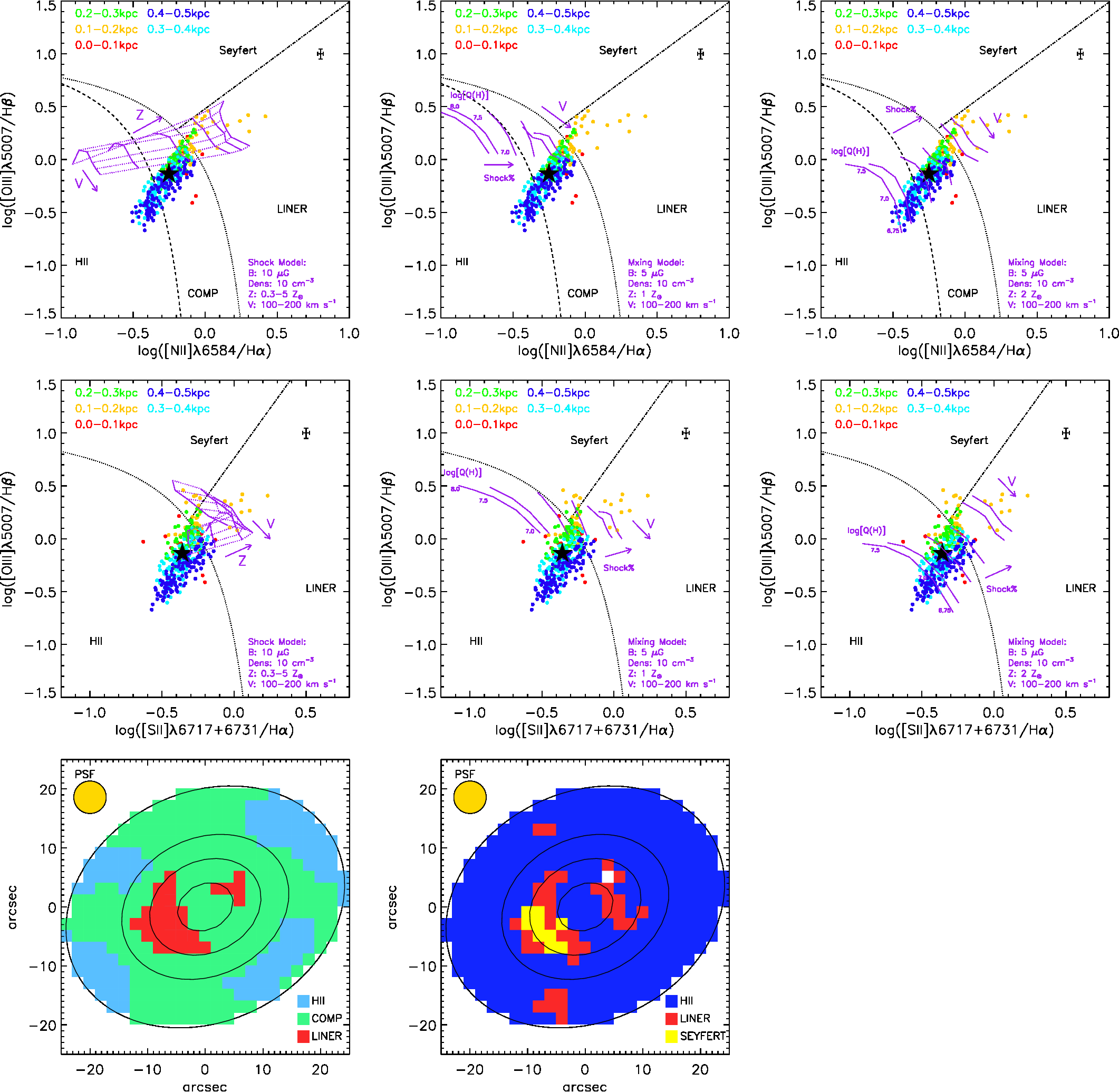}
\caption{Top panel: The BPT diagrams of spectra in the central 
$500\ \mathrm{pc} \times 500\ \mathrm{pc}$ region. 
The dots represent the line ratios of different spxels, and the colors 
reflect their galactocentric radii. The filled black stars show the 
integrated line ratios across this region. Dashed and dotted curves show 
the AGN/star-formation selection criteria of \citet{Kauffmann2003} and 
\citet{Kewley2001}, respectively. The dashed dotted lines show the 
Seyferts/LINERs demarcation of \citet{CidFernandes2010} and \citet{Kewley2006} 
in the [OIII]$\lambda$5007/H$\beta$ vs [NII]$\lambda$6584/H$\alpha$ diagram and the 
[OIII]$\lambda$5007/H$\beta$ vs [SII]$\lambda$6716+$\lambda$6731/H$\alpha$ diagram, 
respectively. The purple curves present the slow shock models from \citet{Rich2010} 
and \citet{Rich2011}. The detailed parameters of these models are introduced in 
\S4.1.3. Bottom panel: The schematic maps of the 
classifications based on the BPT diagrams.}
\label{fig-bpt}
\end{center}
\end{figure}

The ``Baldwin, Phillips, $\&$ Terievich''(BPT) diagrams 
\citep{Baldwin1981,Veilleux1987} of spectra in the central 
$500\ \mathrm{pc} \times 500\ \mathrm{pc}$ region are shown in 
the top panel of Figure \ref{fig-bpt}. We ignore the data 
points with emission lines in the corresponding diagram having 
$\mathrm{S/N} < 5$. The data points are color coded with different 
galactocentric radii. In the bottom panel of Figure \ref{fig-bpt}, 
we present the schematic maps of classifications based on the BPT diagrams. 
In the [OIII]$\lambda$5007/H$\beta$ vs [NII]$\lambda$6584/H$\alpha$ 
diagram, regions with nebular emission dominated by the HII regions 
only lie at galactocentric radius larger than $300\ \mathrm{pc}$, while most 
regions within the $300\ \mathrm{pc}$ are located in the composite part. The 
regions associated with the CRIGS are mainly located at the LINER 
part. In the [OIII]$\lambda$5007/H$\beta$ vs 
[SII]$\lambda$6716+$\lambda$6731/H$\alpha$ diagram, most regions 
are located in the star-forming part, although a large fraction of 
them are located in the composite part of the [OIII]$\lambda$5007/H$\beta$ 
vs [NII]$\lambda$6584/H$\alpha$ diagram. The regions associated with 
the CRIGS are also located in the Seyfert/LINER region.  

As shown in the classification schematic map, most regions in the 
CRIGS present LINER-like emission. In the rest of this paper, we 
aim at characterizing the nature of this structure through analysing 
the excitation mechanism of the LINER-like emission (\S4.1) and 
quantifying the kinematics properties of the ionized gas (\S4.2) 
in the CRIGS.   

\section{Analysis} 

\subsection{Excitation Mechanism of the CRIGS in NGC 1042}

Several ionization mechanisms can give rise to a LINER-like emission. 
The dominant mechanism can vary in different galaxies and even in 
different regions within the same galaxy. AGNs, post-asymptotic giant 
branch (p-AGB) stars and shocks are the main candidates. \citep{Ho2008}. 
In this section, we discuss the dominant ionization mechanism of the 
CRIGS in NGC 1042.

\subsubsection{Photoionization from the Central AGN}

Photoionization from the central AGNs has been considered as the 
most important ionization mechanism of the LINER-like emission in the 
central few hundred pc of galaxies \citep{Ferland1983,Halpern1983,
Pequignot1984,Binette1985,Ho1993,Groves2004}. We investigate the role 
of this mechanism in the CRIGS through analysing the distribution of 
the ionization parameter as a function of radius in the central 
$500\ \mathrm{pc} \times 500\ \mathrm{pc}$ region of NGC 1042.

The dimensionless ionization parameter has been widely used to describe 
the ionization state of gas in galaxies. This parameter is defined as 
$\mathrm{U} = \mathrm{Q}/(\mathrm{n}_{e}\mathrm{c})$, where Q is the flux 
density of the ionizing photons passing through the gas clouds, n$_{e}$ 
is the electric density of the ionized gas, and c is the speed of light. 
Due to the central ionizing property of AGNs, one can predict an r$^2$ 
dilution of Q around this ionization source. When combining with the 
observed n$_{e}$ of the ionized gas, we can estimate the behaviour of U 
as a function of radius if the gas is photoionized by AGNs.

\begin{figure}[htbp]
\begin{center}
\includegraphics[width=0.5\textwidth]{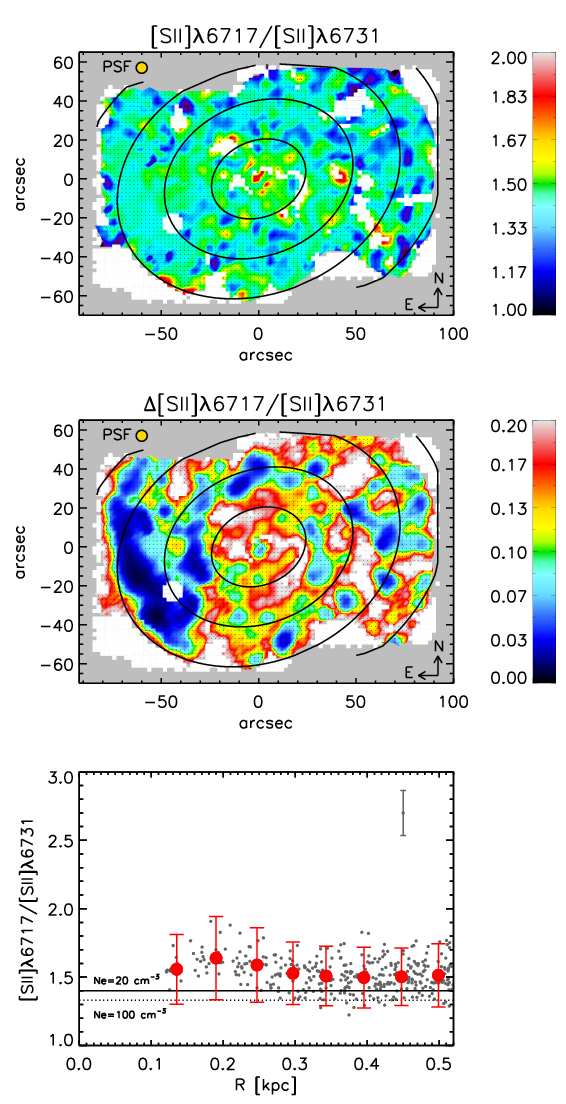}
\caption{Top panel: The emission-line ratio map of 
[SII]$\lambda$6716/[SII]$\lambda$6731 in NGC 1042. 
Middle panel: The map of relative error of 
[SII]$\lambda$6716/[SII]$\lambda$6731 in NGC 1042. 
In these two figures, the spaxels contaminated by foreground 
stars and having line S/N below 5 are masked. The gold circles 
in the top left corners present the $5.6\arcsec$ FWHM PSF of 
the VENGA data. The black contours represent the constant radii 
in steps of $0.5\ \mathrm{kpc}$. Bottom panel: The radial profile 
of the [S II] ratio in the central $500\ \mathrm{pc}$ scale. The 
spaxels having line S/N below 5 are not included. The [S II] ratios 
within the central $120\ \mathrm{pc}$ region are not shown due to 
the limit of the FWHM of PSF (5.6$\arcsec$) in our observation. Gray 
dots present the value of the [S II] ratios in each spaxel. Gray error 
bars in the top right corner show the median error of the [S II] ratios. 
Red dots show the median of the [S II] ratios within the bin of 
$50\ \mathrm{pc}$. The solid and dotted lines correspond to 
$\mathrm{n}_{e} = 20\ \mathrm{cm}^{-3}$ and 
$100\ \mathrm{cm}^{-3}$, respectively. }
\label{fig-density}
\end{center}
\end{figure}

Figure \ref{fig-density} shows the maps of 
[SII]$\lambda$6716/[SII]$\lambda$6731 and its error, 
and the radial profile of [SII]$\lambda$6716/[SII]$\lambda$6731 in the central 
$500\ \mathrm{pc}$ scale. The [SII]$\lambda$6716/[SII]$\lambda$6731 is a function 
of n$_{e}$ and only weakly depends on the gas temperature. It has been widely 
used as an indicator for n$_{e}$. As shown in Figure \ref{fig-density}, the line 
ratio of the [SII] doublets indicates a fairly constant n$_{e}$ as a function of 
radius in the central $500\ \mathrm{pc}$. Thus, if the clouds are photoionized by 
the central AGN, the predicted behaviour of U in this region should in principle 
follow that of Q and also shows an $\mathrm{r}^2$ dilution. This argument assumes 
a constant gas filling factor throughout the inner few hundred pc of the galaxy. 
An increased filling factor could result in a slope shallower than 2, but we 
should still expect a continuous decrease in U as a function of radius.

The observed distribution of U, which can be indicated 
by the [OIII]$\lambda$5007/H$\beta$ (Figure \ref{fig-lineratiozoom} (a)), are 
inconsistent with this prediction. Based on the photoionization models, \citet{Yan2012} 
found that the [OIII]$\lambda$5007/[SII]$\lambda$6716+6731 ratio can also be used to 
trace U. Thus, we also present in Figure 
\ref{fig-lineratiozoom} (d) the map of the [OIII]$\lambda$5007/[SII]$\lambda$6716+6731. 
This line ratio has similar characteristics with the [OIII]$\lambda$5007/H$\beta$ 
map, which shows an enhancement in the CRIGS, $\sim 100\ \mathrm{pc}$-$300\ \mathrm{pc}$ 
away from the nucleus. Therefore, from a morphological point of view, photoionization 
from the central AGN can not be the mechanism of the observed LINER-like emission 
in the CRIGS.

\subsubsection{post-AGB Stars}

Photoionization from post-AGB stars is another mechanism for the 
LINER-like emission in galaxies. These stars can 
produce a diffuse radiation field and provide sufficient ionizing 
photons to account for the extended LINER-like emission in galaxies 
\citep{Binette1994,Stasinska2008,Sarzi2010,CidFernandes2011,Yan2012,
Singh2013}. 

\begin{figure}[htbp]
\begin{center}
\includegraphics[width=\textwidth]{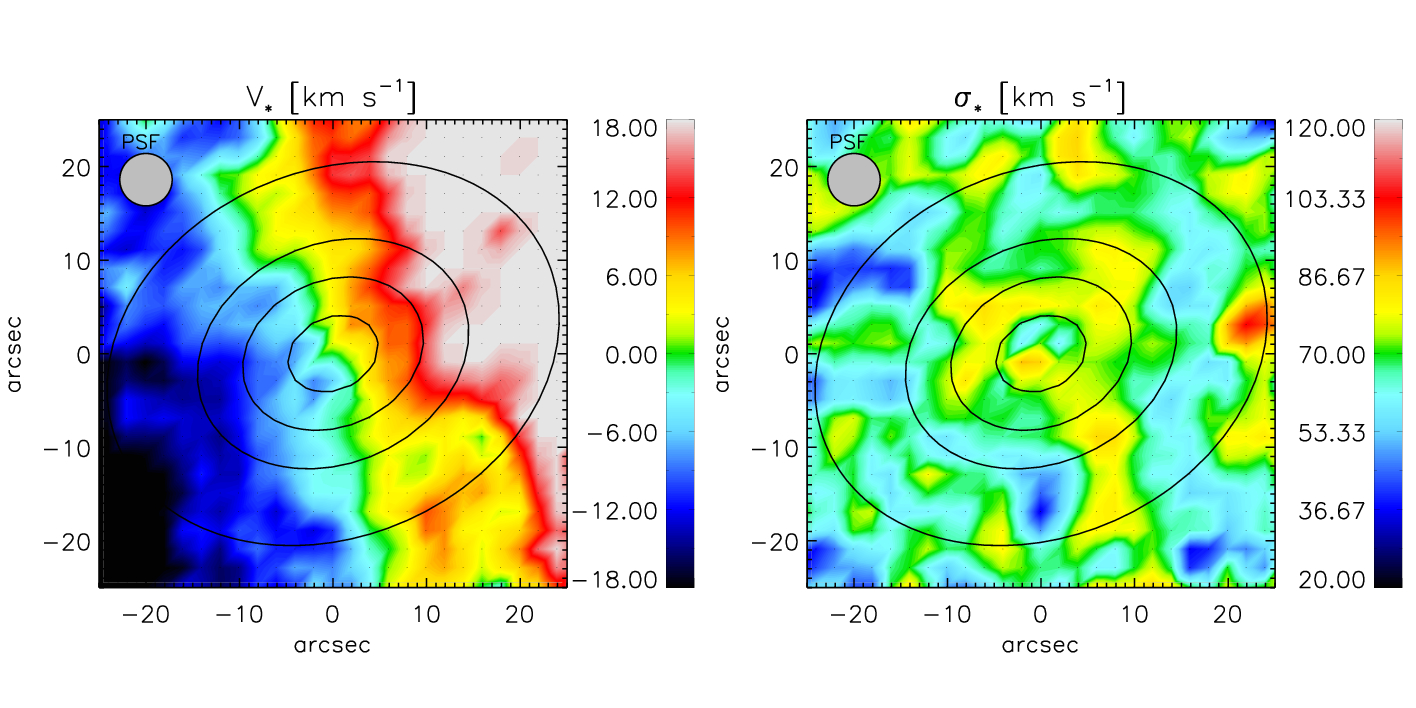}
\caption{The zoomed-in maps of the velocity field and the velocity 
dispersion of the stars in the central 
$500\ \mathrm{pc} \times 500\ \mathrm{pc}$ of NGC 1042. The gray 
circles in the top left corners present the $5.6''$ FWHM PSF of 
the VENGA data. The black contours represent the constant radii 
of 0.1, 0.2, 0.3 and $0.5\ \mathrm{kpc}$.}
\label{fig-vstar}
\end{center}
\end{figure}

However, we do not think that the post-AGB stars are the dominant 
mechanism behind the observed nebular emission in the CRIGS. Post-AGB 
star is the final phase of the stellar evolution for stars with low and 
intermediate initial masses \citep{vanWinckel2003}. Therefore, the 
spatial distribution of post-AGB stars should follow that of the overall 
stellar distribution and present an extended smooth ionization field over 
the whole galaxy. This is inconsistent with the morphology of the CRIGS. 
In addition, as shown in Figure \ref{fig-vstar}, the stars follow the 
regular circular rotation in the CRIGS, the disturbed motion and enhanced 
velocity dispersion only appear in the gas. This suggests that the 
underlying ionization mechanism in the CRIGS is strongly connected with 
the gas, not with the stars.
 
\subsubsection{Shock Excitation}

Radiative shocks can also produce the LINER-like emissions in galaxies. 
The collisional excitation in the post-shock region radiates a large 
amount of photons and can generate the LINER-like emission in galaxies.
The shock-induced LINER-like emission has been observed in systems 
associated with the galactic winds \citep{Veilleux2003,Monreal-Ibero2006,
Sharp2010,Rich2010,Rich2011,Kehrig2012,Ho2014,Arribas2014} and IGM 
in clusters \citep{Farage2010,McDonald2012}. 

We compare the observed properties of the ionized gas (line ratios and 
velocity dispersion) in the CRIGS with the predictions of the radiative 
shock models. The models of steady-flow radiative shocks have been developed 
by \citet{Sutherland1993,Dopita1995,Dopita1996} and summarized in 
\citet{Dopita2003}. In these models, the radiative fluxes and emission-line 
spectra of different shock components are calculated under four physical 
parameters: preshock density, shock velocity, magnetic parameter and medium 
metallicity. The model-predicted flux ratios of different diagnostic lines 
and the corresponding shock velocities can be used to compare with the 
observed emission line ratios and velocity dispersion of gas. Of course, 
the shock-induced enhancement of gas velocity dispersion strongly depends 
on the shock geometry and we do not expected a one-to-one match. But 
considering that the shocks can always accelerate the medium and the 
shock front can not be fully resolved within the observed spatial 
resolution, some positive correlation between the shock velocity and the 
velocity dispersion of shocked gas is excepted. This kind of correlation 
has indeed been observed in Micro-quasar S26 \citep{Dopita2012}. 

In the fast radiative shock models 
($\mathrm{V}_{s} >\ \sim 200\ \mathrm{km}\ \mathrm{s}^{-1}$), 
the ionizing photos from the post-shock region will go through the preshock 
medium and generate a photoionization front (precursor) ahead the shock 
front, which is also included in the calculation of emission line spectrum. 
In the slow radiative shock models 
($\mathrm{V}_{s} <\ \sim 200\ \mathrm{km}\ \mathrm{s}^{-1}$), 
the ionizing photos from the post-shock region will be absorbed in the 
vicinity of the shock front. The radiation contribution of photoionization 
precursor is negligible and the detailed emission line spectrum is more 
related to the shock itself. Since the velocity dispersion of the ionized 
gas in the CRIGS ($90\textrm{-}130\ \mathrm{km}\ \mathrm{s}^{-1}$) is 
significantly lower than the typical velocities of the fast radiative 
shock models, we only compare our observations with the slow radiative 
shock models.  

In recent years, a series of slow radiative shock models have been 
developed and used to test the shock excitation in different environments 
\citep{Farage2010,Rich2010,Rich2011,McDonald2012,Ho2014}. In \citet{Rich2010}, 
the model grids are calculated for fully pre-ionized medium with cloud densities 
$\mathrm{n} = 10\ \mathrm{cm}^{-3}$, metallicities 
$\mathrm{Z} = 0.3,\ 0.5,\ 1,\ 2,\ 3$ and $5\ \mathrm{Z}_{\odot}$, shock 
velocities range from 100 to $200\ \mathrm{km}\ \mathrm{s}^{-1}$ 
(in steps of $20\ \mathrm{km}\ \mathrm{s}^{-1}$) and $\mathrm{B} = 5\ \mu\mathrm{G}$. 
In \citet{Rich2011}, the slow radiative shock model is combined with the 
photoionization model of HII regions to produce a mixing model, in which 
the line ratios vary with the fraction of the shock components. The model 
grids of the HII region are generated with varying ionization parameters 
and metallicities. For the model grids with $\mathrm{Z} = 1\ \mathrm{Z}_{\odot}$ 
and $2\ \mathrm{Z}_{\odot}$, log[Q(H)] is from $7.0\ \mathrm{cm}\ \mathrm{s}^{-1}$ 
to $8.0\ \mathrm{cm}\ \mathrm{s}^{-1}$ and from $6.75\ \mathrm{cm}\ \mathrm{s}^{-1}$ 
to $7.5\ \mathrm{cm}\ \mathrm{s}^{-1}$, respectively.

We plot our observed BPT diagrams together with the models from 
\citet{Rich2010,Rich2011} in Figure \ref{fig-bpt}. In general, these models can 
explain some of the data points in the CRIGS. The models in the 
[OIII]$\lambda$5007/H$\beta$ vs [SII]$\lambda$6716+$\lambda$6731/H$\alpha$ diagram 
can provide a better match to the data points in the CRIGS than that in 
the [OIII]$\lambda$5007/H$\beta$ vs [NII]$\lambda$6584/H$\alpha$ diagram. In the 
mixing models, the grids with higher shock fractions can match the data better. 
The matched shock velocities in these models range from 
$100\ \mathrm{km}\ \mathrm{s}^{-1}$ to $180\ \mathrm{km}\ \mathrm{s}^{-1}$, which 
is roughly consistent with the velocity dispersion of the ionized gas in the CRIGS 
($90\textrm{-}130\ \mathrm{km}\ \mathrm{s}^{-1}$). Based on the above comparisons, 
we propose that the enhanced emission-line ratios and velocity dispersion in the 
CRIGS are related to the shock phenomenon.

\subsection{Kinematic Analysis of the Ionized Gas}

Radiative shocks are strongly related to the kinematic behaviour 
of the ionized gas. Inflows induced by the non-asymmetric structures, 
outflows driven by the starbursts or AGNs and cloud collisions can 
produce shocks in the ISM. The kinetic energy of gas supersonic 
motions can be effectively dissipated through radiative shocks. 
In order to understand the kinematic properties of the ionized gas 
in the CRIGS, we compare the observed velocity field of the ionized 
gas with the velocity field model of a pure rotating disk.

\begin{figure}[htbp]
\begin{center}
\includegraphics[width=\textwidth]{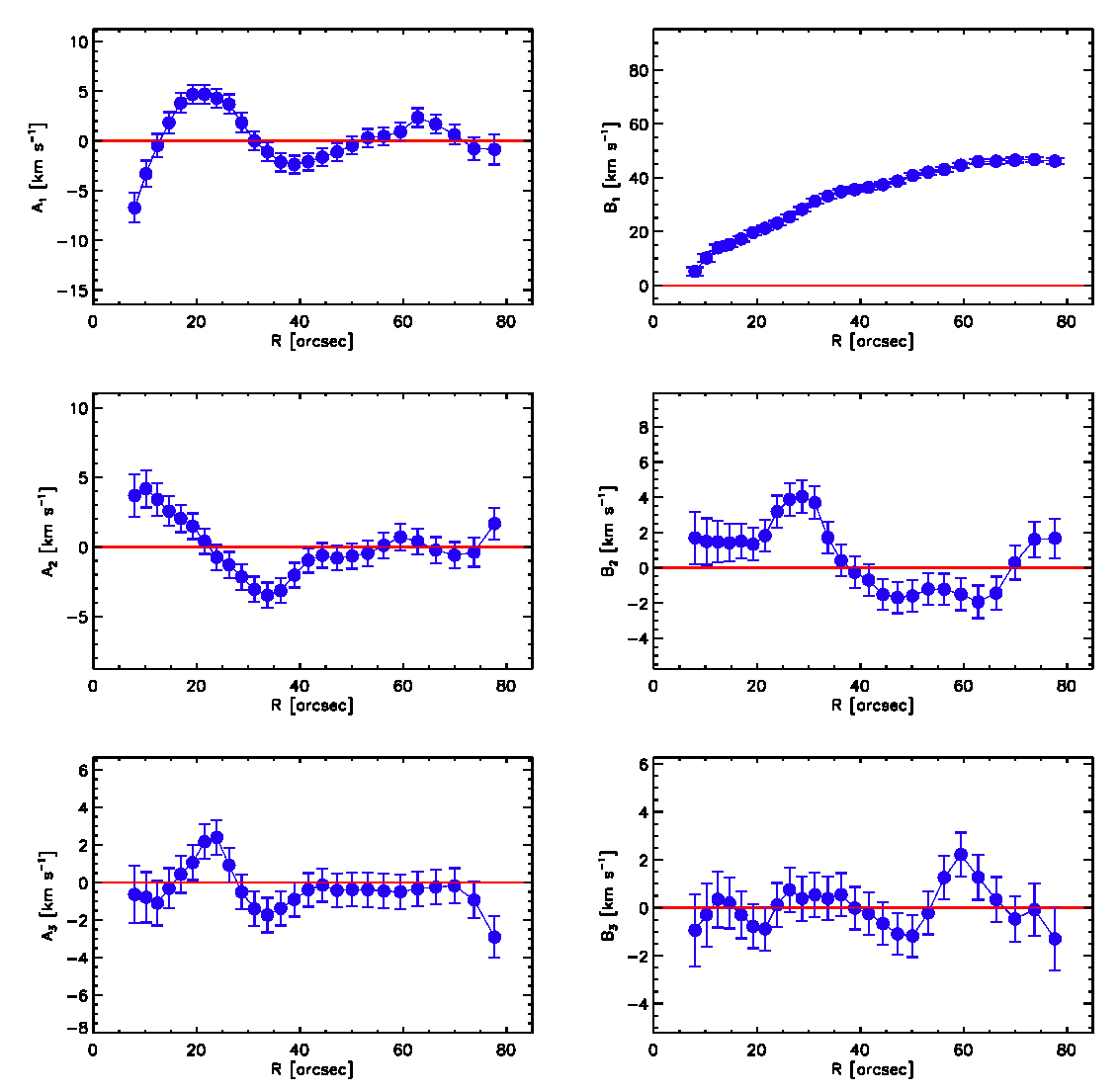}
\caption{The radial profiles of the coefficients of different Fourier 
components obtained from the harmonic decomposition modelling. The range 
of the elliptical major axis in the harmonic decomposition modelling is 
from $6\arcsec$ to $80\arcsec$, corresponding to the physical scale from 
$120\ \mathrm{pc}$ to $1.6\ \mathrm{kpc}$. The minimum radius is set to be 
slightly larger than the FWHM of the PSF ($5.6\arcsec$) in our observation. 
The outermost radius is set by the requirement that there are at least 
$75\%$ of the data available in the elliptical annuli.}
\label{fig-vfourier}
\end{center}
\end{figure}
 
We construct the velocity field model using the method of harmonic 
decomposition modelling. This method uses different Fourier components 
of the line-of-sight (LOS) velocity to describe the kinematic properties 
of galaxies and construct the corresponding velocity field models (A more 
detailed description of the harmonic decomposition modelling is in Appendix A). 
We use the IDL package KINEMETRY \citep{Krajnovic2006} to perform the harmonic 
decomposition modelling. First, we determine the geometry parameters of the pure 
rotating disk and the systemic velocity of NGC 1042 (see Appendix B for details). 
Next, we fix these parameters and decomposed the LOS velocity field into a 
series of Fourier components in a set of elliptical annuli along the semi-major 
axis of the galaxy. We expand the Fourier terms to the third order 
($V_{los}(R,\psi) = A_0(R)+\sum\limits_{n=1}^3[A_n(R){\rm sin}(n\psi)+B_n(R){\rm cos}(n\psi)]$) 
and show the radial profiles of their coefficients in 
Figure \ref{fig-vfourier}. As the term $B_1(R){\rm cos}(\psi)$ describe 
the rotational velocity component of the LOS velocity, we use it to build 
the velocity field model of the pure rotating disk. The range of the 
elliptical major axis is from $6\arcsec$ to $80\arcsec$, corresponding to 
the physical scale from $120\ \mathrm{pc}$ to $1.6\ \mathrm{kpc}$. The 
minimum radius is set to be slightly larger than the FWHM of the PSF 
($5.6\arcsec$) in our observation. The outermost radius is set by the 
requirement that there are at least $75\%$ of the data available in the 
elliptical annuli.

\begin{figure}[htbp]
\begin{center}
\includegraphics[width=\textwidth]{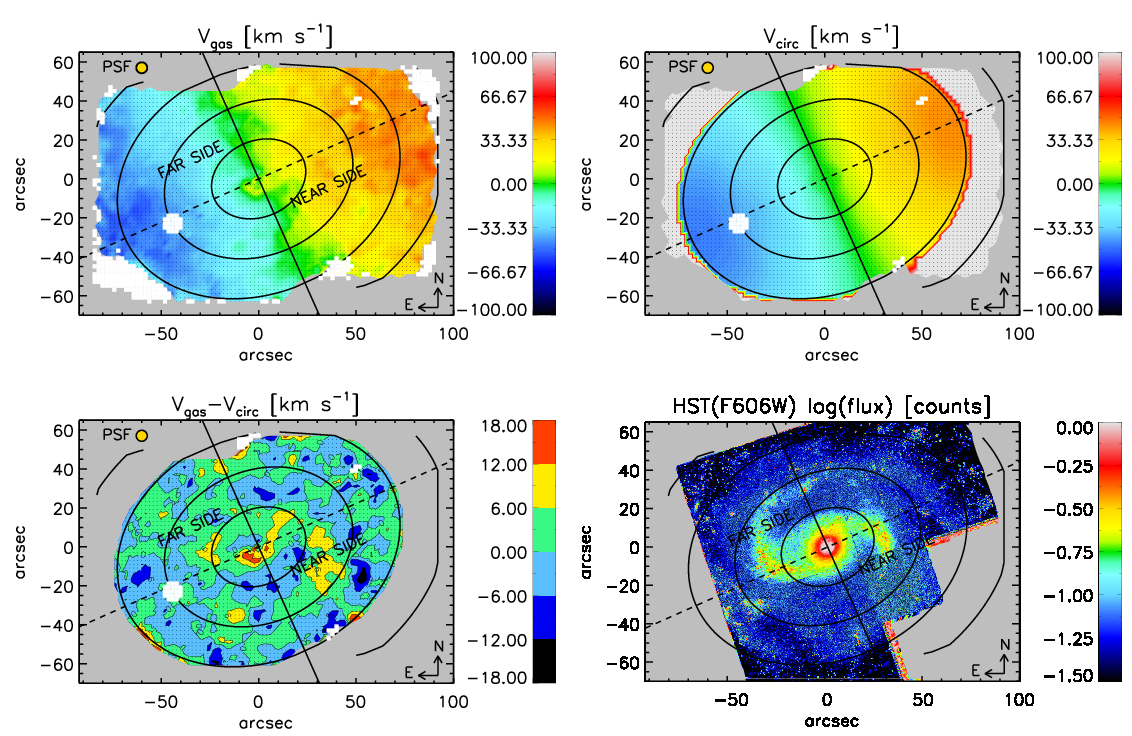}
\caption{The observed velocity field of the ionized gas, the velocity field 
model of the pure rotating disk, the residual velocity field (subtract model 
from the observed map) and the HST 606W image of NGC 1042. The kinematic major 
axis and minor axis are shown as black dashed and solid line, respectively. 
The gold circles in the top left corners present the $5.6\arcsec$ FWHM PSF 
of the VENGA data. The black contours represent the constant radii in 
steps of $0.5\ \mathrm{kpc}$.}
\label{fig-vgasmodel}
\end{center}
\end{figure}

The observed velocity field of the ionized gas, the velocity field model of 
the pure rotating disk, the residual velocity field (subtract model from the 
observed map) and the HST 606W image of NGC 1042 are shown in Figure \ref{fig-vgasmodel}. 
The minimum error of our velocity measurements is about $6\ \mathrm{km}\ \mathrm{s}^{-1}$. 
Thus, we show the residual velocity field with velocity bin of 
$6\ \mathrm{km}\ \mathrm{s}^{-1}$. Assuming the spiral arms are trailing and 
considering the rotation direction of the velocity field, we determine the 
near side and far side of this galaxy to be Southwest (SW) and Northeast (NE), 
respectively. The typical value of the residual velocity ranges from 0 to 
$\sim 6\ \mathrm{km}\ \mathrm{s}^{-1}$, which is within the error bars of the 
LOS velocity. For the bright star-forming regions in the spiral arms, the residual 
velocity can enhance to $\sim 12\ \mathrm{km}\ \mathrm{s}^{-1}$. This could be the 
result of supersonic turbulence maintained by the energy output through supernovae 
and stellar winds \citep{Scalo2004,Sellwood2014}. However, we notice that only 
part of the bright star-forming regions are associated with the enhancement of 
the residual velocity.

\begin{figure}[htbp]
\begin{center}
\includegraphics[width=\textwidth]{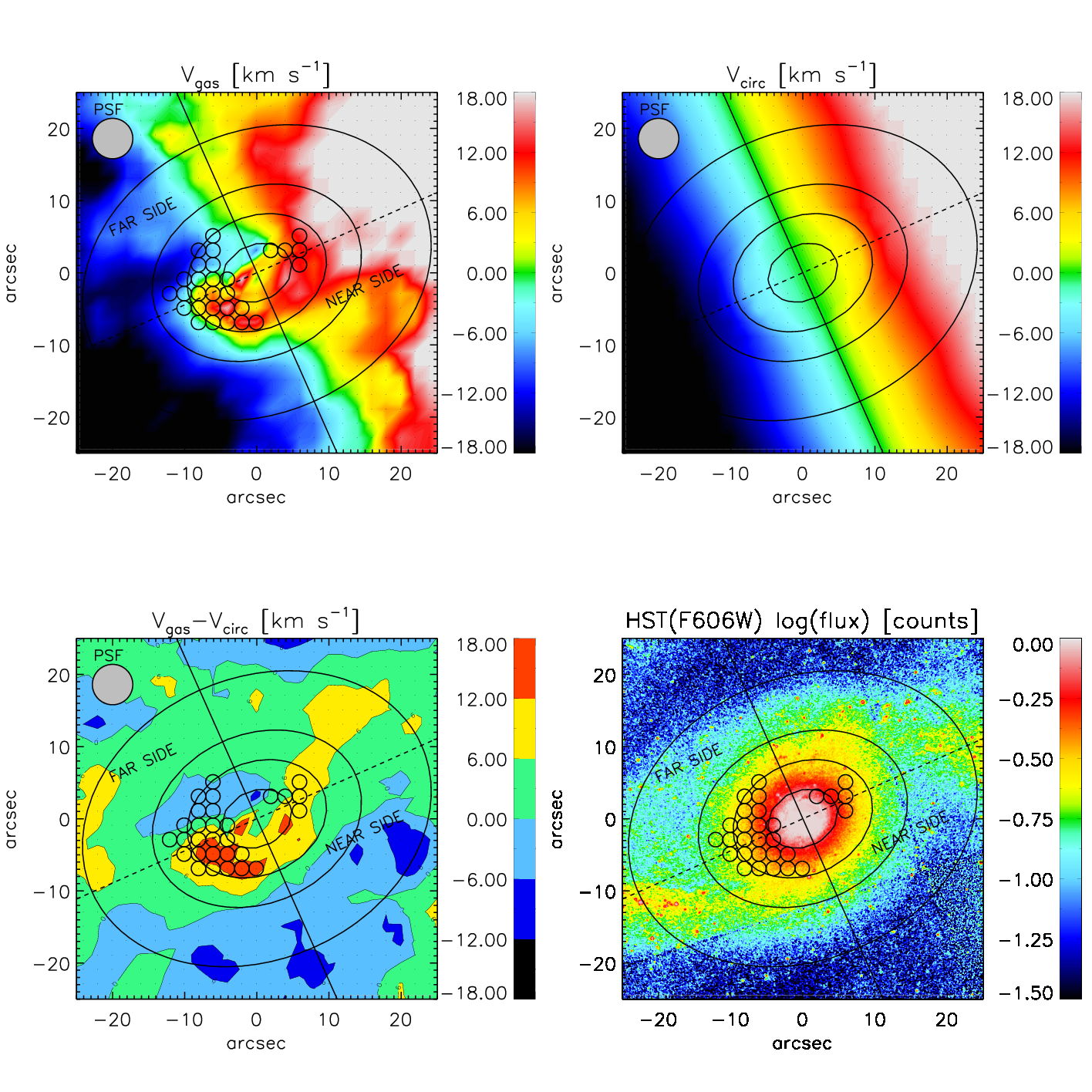}
\caption{The observed velocity field of the ionized gas, the velocity field 
model of the pure rotating disk, the residual velocity field (subtract model 
from the observed map) and the HST 606W image of the central 
$500\ \mathrm{pc} \times 500\ \mathrm{pc}$ region in 
NGC 1042. The kinematic major axis and minor axis are shown as black dashed 
and solid line, respectively. Small black circles mark the CRIGS. The gray 
circles in the top left corners present the $5.6\arcsec$ FWHM PSF of the VENGA 
data. The black contours represent the constant radii of 0.1, 0.2, 0.3 and 
$0.5\ \mathrm{kpc}$. }
\label{fig-vgasmodelzoom}
\end{center}
\end{figure}

As shown in Figure \ref{fig-vgasmodelzoom}, in the central 
$500\ \mathrm{pc} \times 500\ \mathrm{pc}$ region, we find that the residual 
velocity associated with the CRIGS has a significant enhancement and reach to 
the largest value in the residual velocity field 
($\sim 20\ \mathrm{km}\ \mathrm{s}^{-1}$), 
indicating strong non-circular motions in this structure.
On the near side we see redshifted velocities and on the NE quadrant of the 
far side we see blueshifted emission. This amounts for a clear signature of 
ionized gas inflow associated with the CRIGS. Further evidences come from 
the radial profiles of different Fourier coefficients obtained from the 
harmonic decomposition modelling. As we introduced in Appendix A, the first-order 
Fourier terms ($A_1{\rm sin}(\psi)$ and $B_1{\rm cos}(\psi)$) describe the 
radial and rotational velocity components of the LOS velocity. The higher-order 
Fourier terms ($A_n{\rm sin}(n\psi)$ and $B_n{\rm cos}(n\psi)$, $n \geqq 2$) can 
provide the information about the perturbations of the gravitational potential.
As shown in Figure \ref{fig-vfourier}, the radial velocity component ($A_1$) ranges 
from $0\ \mathrm{km}\ \mathrm{s}^{-1}$ to $-7\ \mathrm{km}\ \mathrm{s}^{-1}$ in the 
central 8$\arcsec$-15$\arcsec$ (160 pc-300 pc) region, associated 
with the CRIGS. The negative sign indicates the inflowing feature of this component. 
In this region, the absolute value of this inflowing velocity component is comparable 
to that of the rotational velocity component ($B_1$), ranging from 
$5\ \mathrm{km}\ \mathrm{s}^{-1}$ to $15\ \mathrm{km}\ \mathrm{s}^{-1}$. Although 
the velocity component $A_2$ also have some contribution within the inner 15$\arcsec$ 
($300\ \mathrm{pc}$), which may indicate the $\mathrm{m} = 1$ potential perturbation 
in this region (e.g. lopsided disk), the other higher order velocity components 
($B_2$, $A_3$, $B_3$) only have limited contribution to the LOS velocity. These 
results suggest the ionized gas inflow is the main contributor of the non-circular 
motions in the CRIGS.

Interestingly, the morphology of the inflowing gas is asymmetric. As shown 
in Figure \ref{fig-vgasmodelzoom}, the inflow signature is dominated by the redshifted 
velocities in the near side of NGC 1042. In order to quantify the significance of 
this asymmetry, we compare the inflow timescale (the timescale of the inflowing gas 
moving to the galaxy center) with the dynamical timescales in the central disk (at 
radius of $300\ \mathrm{pc}$ ($R$)). Based on the measured rotation velocity 
($V_{rot} = 24\ \mathrm{km}\ \mathrm{s}^{-1}$), we obtain the dynamical timescale 
$T_{dyn} = 2 \pi R/V_{rot} \sim 7.7 \times 10^{8}\ \mathrm{yr}$. In the meantime, 
assuming a constant inflow velocity ($V_{in} = 32\ \mathrm{km}\ \mathrm{s}^{-1}$, see \S5.1 for details), 
we obtain the inflow timescale $T_{in} = R/V_{in} \sim 9.2 \times 10^{7}\ \mathrm{yr}$. 
$T_{dyn}$ is about 8 times $T_{in}$, suggesting that the gas inflows cannot last 
for one rotation period. Thus, the asymmetry of the inflowing gas is expected as 
the gas does not have time to circle the disk before it is being induced to the center.

In disk galaxies, the gravity torques induced by the non-axisymmetric 
features (such as bars, ovals, and spiral arms) can drive gas inward 
to the central region of galaxies \citep{Shlosman1989,Shlosman1990,
Kormendy2004,Martini2004,Jogee2006}. The inflowing gas is usually shocked 
and can be traced by dust lanes \citep{Athanassoula1992}. As we introduced 
in \S1.1 and shown in Figure \ref{fig-vgasmodel} and \ref{fig-vgasmodelzoom}, 
the inner spiral arms in NGC 1042 are very open and sharply curve towards 
the central region, which can produce a bar-like structure and enhance the 
perturbation strength of the gravitational potential \citep{Buta2005,Buta2009}. 
In Figure \ref{fig-vgasmodelzoom}, we find that the shocked gas is located 
at the trailing side and the end of the inner spiral arms. Thus, we propose 
that this spiral arm structure can produce a similar dynamical effect as 
a bar to drive the radial gas inflow. If two spiral arms are considered as 
the boundary of a ``bar'', we would expect the shocked gas inflow appear 
along the trailing side of the spiral arms and at the point where spiral arms 
end \citep{Li2015}, which is supported by our observation. Previous studies 
have found two excellent examples (NGC 4321 and M 51) of secular evolution 
induced by spiral arms \citep{Kormendy2004a,Kormendy2004}. In these galaxies, 
the regular global spiral arms directly wind down to the galaxy center and 
produce nuclear star formation. Our observational results of NGC 1042 provide 
another example of spiral arm induced gas inflows in disk galaxies.
 
\section{Discussion}

In \S4 we demonstrate that the CRIGS is shock excited and the shock is
produced by the inflow of the ionized gas driven by the inner spiral arms 
in NGC 1042. Here we estimate the mass inflow rate of the ionized gas and 
further discuss its implications for the AGN feeding and star formation 
in the nuclear star cluster.

\subsection{Mass Inflow Rate of the Ionized Gas} 

\begin{figure}[htbp]
\begin{center}
\includegraphics[width=0.55\textwidth]{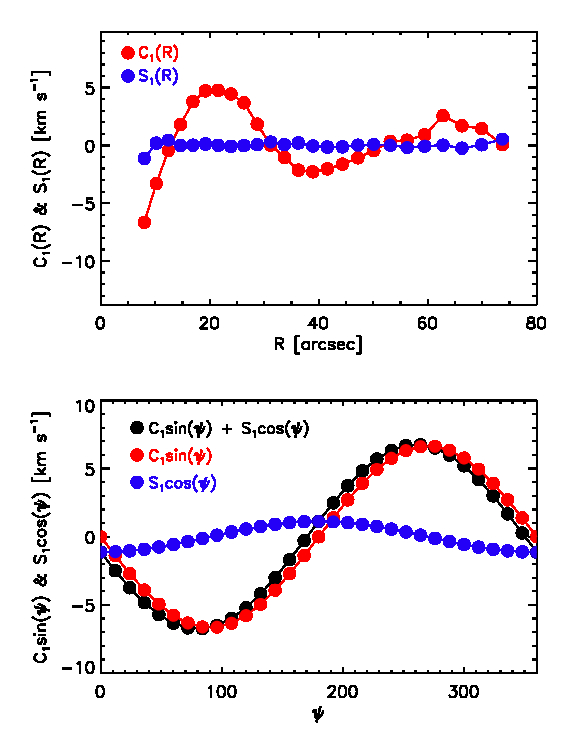}
\caption{Top panel: The radial profiles of the coefficients of 
$C_1(R)$ and $S_1(R)$ in the residual velocity. The range of the 
elliptical major axis in the decomposition process is from $6\arcsec$ to 
$75\arcsec$, corresponding to the physical scale from $120\ \mathrm{pc}$ 
to $1.5\ \mathrm{kpc}$. The minimum radius is set to be slightly larger 
than the FWHM of the PSF ($5.6\arcsec$) in our observation. The outermost 
radius is set by the requirement that there are at least $75\%$ of the 
data available in the elliptical annuli. Bottom panel: The azimuthal 
distributions of the Fourier components $C_1{\rm sin}(\psi)$ and 
$S_1{\rm cos}(\psi)$ at the annulus with radius of $150\ \mathrm{pc}$ 
(the median radius of the CRIGS). The error bars are smaller than the 
data symbols shown in the figure, thus are not visible.}
\label{fig-vresfourier}
\end{center}
\end{figure}

As shown in \S4.2.3, the observed LOS velocity in the CRIGS is mainly
dominated by the inflowing velocity component ($A_1$ term). In order to 
quantify this and separate the radial flow (which points towards the center, 
and may feed the nuclear activities) and the azimuthal streaming flow (which 
may just circle around the CRIGS, and not contribute to the feeding of 
the nuclear activities), we decompose the residual velocity field shown 
in Figure \ref{fig-vgasmodel} into two Fourier terms, as 
$V_{res}(R,\psi) = C_1(R){\rm sin}(\psi) + S_1(R){\rm cos}(\psi)$.
The components $C_1(R){\rm sin}(\psi)$ and $S_1(R){\rm cos}(\psi)$
correspond to the radial flow component and the azimuthal streaming component, 
respectively. We again use the Kinemetry software to execute the decomposition 
and show the results in Figure \ref{fig-vresfourier}. The top panel shows the 
$C_1$ and $S_1$ as a function of radius. It shows that the radial inflow 
dominates the residual velocity and the azimuthal streaming flow is negligible. 
In the bottom panel of Figure \ref{fig-vresfourier}, we also show the azimuthal 
distribution of the components $C_1{\rm sin}(\psi)$, $S_1{\rm cos}(\psi)$ and 
their sum at the annulus with radius of $150\ \mathrm{pc}$ (the median radius 
of the CRIGS). In the whole annulus, the azimuthal streaming flow is negligible. 
Thus, we use the whole residual velocity ($\sim 20\pm6\ \mathrm{km}\ \mathrm{s}^{-1}$) 
associated with the CRIGS to approximate the inflow velocity along the line of sight. 
By adapting the inclination angle of $38.7^{\circ}$, we obtain the deprojected 
inflow velocity as ${V}_{in} \sim 32\pm10\ \mathrm{km}\ \mathrm{s}^{-1}$.

Having estimated the inflow velocity, we now estimate the mass inflow 
rate of the ionized gas. Since the CRIGS has a ring-like morphology, we 
assume that the inflowing gas is ring-like at radius $r$ and with height 
$h$, and we calculate the integral of the flux of matter through a 
cylindric area of radius $r$ and height $h$. Thus the mass inflow 
rate can be described as below:
 
\begin{equation}
 \dot{M}_{in}\, = \,2\,\pi\,r\,h\,n_{e}\,m_{p}\,f\,V_{in}\
\end{equation}

\noindent
where $n_{e}$ is the electron density, $m_{p}$ is the mass of the proton, 
$f$ is the filling factor of the ionize gas, and $V_{in}$ is the deprojected 
inflow velocity.
   
$r$ and $h$ can be estimated based on the position and size of the inflow 
region. We use the outer boundary of the CRIGS as the border of the inflow 
region, thus $r$ is about $300\ \mathrm{pc}$. The inflow region extends from 
$100\ \mathrm{pc}$ to $300\ \mathrm{pc}$ from the galaxy center, thus the size of 
this region is $\Delta r \sim 200\ \mathrm{pc}$. Assuming the value of $h$ has 
similar order of magnitude as that of the radial span of the inflow region, 
and considering the projection effect, $h$ should be $\Delta r \times \mathrm{sin(i)}$, 
which gives $\sim 70\ \mathrm{pc}$. Even though this is an assumption, $h$ can not be 
larger than this value, otherwise, the inflow region would appear wider. Therefore, 
$\Delta r \times \mathrm{sin(i)}$ can be considered as an upper limit of $h$. 
Adopting an average $n_{e} \sim 10\ \mathrm{cm}^{-3}$ (see Figure \ref{fig-density}) 
and $f = 0.001$ \citep{SchnorrMueller2011}, we obtain the mass inflow rate as 
$\dot{M}_{in} \sim 1.1\pm0.3 \times 10^{-3}\ \mathrm{M}_{\odot}\ \mathrm{yr}^{-1}$. 
The uncertainty of distance of NGC 1042 (describe in \S1.1) can affect the value 
of $r$, $h$, and the estimation of mass inflow rate, thus, we also calculate the mass 
inflow rate corresponding to the distance of NGC 1042 at $8\ \mathrm{Mpc}$, which is 
$\dot{M}_{in} \sim 2.1\pm0.3 \times 10^{-3}\ \mathrm{M}_{\odot}\ \mathrm{yr}^{-1}$.

\subsection{Implication for Feeding the Nuclear Activities}
 
The mass accretion rate ($\dot{M}$) at the last stable orbit of the blackhole 
can be calculated as:

\begin{equation}
 \dot{M} = \frac{L_{bol}}{c^2\eta}
\end{equation}

\noindent
where $c$ is the velocity of light, $L_{bol}$ is the bolometric luminosity 
of AGN, and $\eta$ is the radiation efficiency which describe the efficiency 
of accreted mass energy converted into radiation. $\eta$ depends on the nature 
of different accretion disks and flows. For LINERs, it has been suggested that 
the accretion disk is geometrically thick, and optically thin 
\citep{Nemmen2006,Yuan2007}. This kind of accretion flow is known as RIAF 
\citep[Radiatively Inefficient Accretion Flow;][]{Narayan2005}, and has a 
typical value of $\eta \sim 0.01$. Using the bolometric luminosity of 
$8.0 \times 10^{39}\ \mathrm{erg}\ \mathrm{s}^{-1}$ for NGC 1042 \citep{Shields2008}, 
we obtain the mass accretion rate 
$\dot{M} \sim 1.4 \times 10^{-5}\ \mathrm{M}_{\odot}\ \mathrm{yr}^{-1}$. 

The mass inflow rate described in \S5.1 is about 100 times the mass accretion 
rate. This result suggests that, the inflowing gas is sufficient to feed the 
AGN activity in NGC 1042, and only a small portion (a few percent) may be 
transferred into the center to feed the blackhole. This is in agreement with 
many previous studies. As discussed in \citet{Ho2008,Ho2009}, most LLAGNs 
in the nearby universe reside in a low or quiescent state and undergo the radiative 
inefficient accretion, even though the fuel is plentiful in the central region 
of these galaxies. They show that the local processes (e.g. mass loss from evolved 
stars and Bondi accretion of hot gas) can provide sufficient fuel for nuclear 
activity. Based on the lack of correlation of blackholes with disks and 
pseudobulges, \citet{Kormendy2011} and \citet{Kormendy2013} also proposed that 
the local stochastic processes can feed the growth of blackholes in disk-dominated 
galaxies at redshift $\sim 0$. In addition, \citet{Genzel2010} summarized the 
effective mass accretion rate from a few hundred pc to several schwarzschild radius 
in the Milk Way and proposed that the angular momentum transport of the gas 
surrounding the Galactic center blackhole could be inefficient. 
High-spatial-resolution observations (e.g. Thirty Meter Telescope, Atacama Large 
Millimeter/submillimeter Array) will be important for the study of AGN feeding 
within $100\ \mathrm{pc}$ of galaxies, which can help to understand the relationship 
between the accretion process of balckholes and its surrounding materials.

In addition to the AGN activity, the nuclear star cluster in NGC 1042 has a young 
stellar population component with the age of $\sim 10\ \mathrm{Myr}$ and mass of 
$\sim 7.94 \times 10^3\ \mathrm{M}_{\odot}$. This implies an averaged star formation 
rate of $\sim 7.94 \times 10^{-4}\ \mathrm{M}_{\odot}\ \mathrm{yr}^{-1}$ in this component. 
The mass inflow rate we obtained above is at similar order of magnitude of this star 
formation rate, which means the inflowing gas is sufficient to feed the star formation 
in the nuclear star cluster and make it grow.

\section{Summary}

Using the IFS data from the VENGA survey, we have explored the feeding 
process of the LLAGN in late-type bulgeless galaxy NGC 1042. The large 
spatial coverage of the Mitchell spectrograph 
observation ($3.5\ \mathrm{kpc} \times 2.5\ \mathrm{kpc}$) 
enables us to map the emission of the ionized gas from several kpc down 
to several hundred pc in this galaxy. By studying the excitation and 
kinematic properties of the ionized gas using various emission-line 
properties, we directly identify the radial gas inflows in the central 
$500\ \mathrm{pc} \times 500\ \mathrm{pc}$ of the galaxy.

From the flux and flux ratio maps of emission lines (including 
[OII]$\lambda\lambda$3726,3729, [OIII]$\lambda$5007, H$\alpha$, 
H$\beta$, [NII]$\lambda$6548,6583, and [SII]$\lambda$6717,6731), 
we find a cirumnuclear ring-like structure of ionized gas (which 
we call CRICS) in the central $100\textrm{-}300\ \mathrm{pc}$ region. 
Using the spatically-resovled BPT diagnostics, we find that this structure 
presents the LINER-like emission. By comparing with the prediction 
of shock ionization models, we conclude that shocks are the dominant 
ionization source in this structure. This result is also supported 
by the disturbed kinematics of the ionized gas (distorted velocity 
field and enhance velocity dispersion) in this structure.

The harmonic decomposition modelling is used to analyse the velocity 
field of the ionized gas and quantify the possible non-circular motions 
driven by the gas flows. We do not find significant non-circular motions 
of the ionized gas at large radii of the disk. Strong non-circular motions 
of the ionized gas only exist within the central 
$500\ \mathrm{pc} \times 500\ \mathrm{pc}$, especially at the CRIGS. 
On the near side we see redshifted velocities and on the NE quadrant of the 
far side we see blueshifted emission, indicating the ionized gas in the CRIGS 
is inflowing. Combining with the HST image of NGC 1042, we find that the CRIGS 
takes place at the end of the inner spiral arms. Since the inner spiral arms 
in NGC 1042 sharply curve towards the central region, which produce a bar-like 
structure, we propose that the inner spiral arms of NGC 1042 can produce a similar 
dynamical effect as a bar to drive the shocked ionized gas inflow. 

We calculate the de-projected inflow velocity at 
$\sim 32\pm10\ \mathrm{km}\ \mathrm{s}^{-1}$ and further estimate the mass 
inflow rate at 
$\sim 1.1\pm0.3 \times 10^{-3}\ \mathrm{M}_{\odot}\ \mathrm{yr}^{-1}$. 
The mass inflow rate is about one hundred times the BH's mass accretion rate 
($\sim 1.4 \times 10^{-5}\ \mathrm{M}_{\odot}\ \mathrm{yr}^{-1}$) and at 
similar order of magnitude of the star formation rate in the nuclear star cluster 
($7.94 \times 10^{-4}\ \mathrm{M}_{\odot}\ \mathrm{yr}^{-1}$). It is large enough 
to feed both the nuclear activity and the star formation in the nuclear star cluster. 
The BH's mass accretion rate is significantly less than the mass inflow rate, suggesting 
that only a small portion of these material may be transferred into the center to feed 
the blackhole. Although the exact feeding processes of the nuclear activity need 
to be further explored via the high-spatial-resolution observation within 
the central $100\ \mathrm{pc}$ region, our study highlights that secular 
evolution can be important in late-type unbarred galaxies like NGC 1042.

\acknowledgments

We thank the anonymous referee for his/her careful and helpful comments, 
which improved the quality of the paper immensely. We also thank Ramya 
Sethuram for her help in refining the paper. R.L. and L.H. acknowledge 
support from the National Natural Science Foundation of China under 
grants No. 11473305, by the Strategic Priority Research Program 
``The Emergence of Cosmological Structures'' of Chinese Academy of Sciences, 
Grant No. XDB09030200. G.B. is supported by CONICYT/FONDECYT, Programa de 
Iniciacion, Folio 11150220. S.J. acknowledges support from NSF grant NSF 
AST-1413652 and the National Aeronautics and Space Administration (NASA) 
JPL SURP Program.

\appendix

\section{Harmonic Decomposition Modelling}
Harmonic decomposition modelling (HDM) can decompose the line-of-sight 
(LOS) velocity field of a galaxy into a series of Fourier components 
which are considered as kinematic components with different azimuthal 
symmetry. By matching the kinematic symmetry of the observed velocity 
field, these components can be used to construct a model velocity field. 
This method has been used to study the stellar and gas kinematics in 
late-type galaxies \citep{Binney1978,Begeman1987,Teuben1991,Franx1994,
Schoenmakers1997,Wong2004}, early-type galaxies \citep{Krajnovic2006,
Emsellem2007,Krajnovic2008,Krajnovic2011,Krajnovic2013} and high-z 
merging systems \citep{Shapiro2008,Gonccalves2010,Swinbank2012,
Alaghband-Zadeh2012,Bellocchi2012}. 

The Fourier components in the HDM are obtained in a set of elliptical 
annuli along the semi-major axis of a galaxy. The shapes of these elliptical 
annuli are described by their geometry parameters, including the positions 
of the centers, position angles (P.A.) and inclinations. For each elliptical 
annulus, the LOS velocity is fitted as

\begin{equation}
V_{los}(R,\psi) = A_0(R)+\sum\limits_{n=1}^k[A_n(R){\rm
  sin}(n\psi)+B_n(R){\rm cos}(n\psi)]
\end{equation}
\noindent
where $R$ is the semi-major axis of the elliptical annuli, and $\psi$ is 
the eccentric anomaly angle. $A_0$, $A_n$ and $B_n$ are the 
corresponding coefficients of different Fourier components. The zero-order 
term ($A_0$) is the systemic velocity of the galaxy. The first-order 
terms ($A_1{\rm sin}(\psi)$ and $B_1{\rm cos}(\psi)$) describe the 
radial and rotational velocity components. The higher-order terms 
($A_n{\rm sin}(n\psi)$ and $B_n{\rm cos}(n\psi)$, $n \geqq 2$) can 
provide information about the perturbations of gravitational potential 
\citep{Schoenmakers1997,Wong2004}. 

In disk galaxies, we assume the gas locate on the galaxy plane and 
ignore their motions in the vertical direction. The LOS velocity 
along each elliptical annulus of a pure rotating disk can be described as: 
$V_{los}(R,\psi) = A_0 + B_1(R){\rm cos}(\psi)$.
In this case, the geometry parameters of elliptical annuli and the systemic 
velocity do not change with $R$. The other Fourier components in Equation 
(1) describe the departures from circular rotation, which can be used to 
quantify different gas flows in disk galaxies. 

\section{Determining Disk Parameters}
In general, the kinematic center is consistent with the photometric 
center in galaxies \citep{Trachternach2008,Haan2008,Neumayer2011,Andersen2013}. 
Considering the effective PSF FWHM of $5.6\arcsec$ in our data cube, we 
assume the kinematic center and photometric center are in agreement 
and use the photometric center from SDSS DR8 \citep{Aihara2011} as 
the center coordinates ($02^{h}40^{m}24^{s}.967$, $-08^{\circ}26\arcmin00\arcsec.76$) 
of NGC 1042. This parameter will be fixed in the subsequent analysis.  

\begin{figure}[htbp]
\begin{center}
\includegraphics[width=0.55\textwidth]{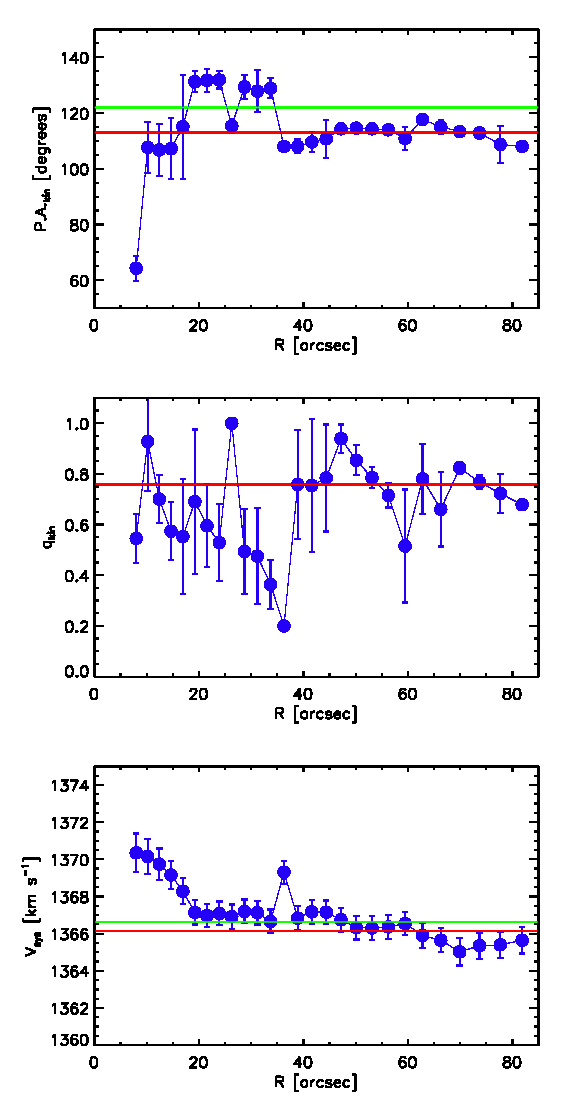}
\caption{The radial profile of the kinematic P.A., the flattening 
and the systemic velocity of NGC 1042 obtained from Kinemetry. The 
range of the elliptical major axis in Kinemetry is from $6\arcsec$ 
to $80\arcsec$, corresponding to the physical scale from 
$120\ \mathrm{pc}$ to $1.6\ \mathrm{kpc}$. The minimum radius is set 
to be slightly larger than the FWHM of the PSF ($5.6\arcsec$) in our 
observation. The outermost radius is set by the requirement that there 
are at least $75\%$ of the data available in the elliptical annuli. 
The red solid lines show the average values of these parameters within 
$40\arcsec < \mathrm{R} < 80\arcsec$. The green solid lines present the 
kinematic P.A. and the systemic velocity obtained from FIT\_KINEMATIC\_PA.}
\label{fig-vparameter}
\end{center}
\end{figure}

In order to determine the systemic velocity, position angle (P.A.) and 
inclination of NGC 1042, we first use Kinemetry \citep{Krajnovic2006} 
to quantify the observed velocity field of the ionized gas and present 
the radial profiles of these parameters. As shown in Figure 
\ref{fig-vparameter}, all parameters have significant 
variances within the inner $40\arcsec$, which could be caused by the 
kinematic twist due to gas flow in this region and the existence of 
the inner spiral arms. While in the outer $40\arcsec < \mathrm{R} < 80\arcsec$, 
these parameters show a relatively flat profile and the kinematics also 
presents the signature of regular circular motion. Therefore, we determine 
the systemic velocity, P.A. and inclination of NGC 1042 to be the radial 
average of these parameters within $40\arcsec < \mathrm{R} < 80\arcsec$. 
The obtained systemic velocity, kinematic P.A. and flattening are 
$1366.3\pm3\ \mathrm{km}\ \mathrm{s}^{-1}$, $114.0\pm1^{\circ}$ and 
$0.78\pm0.5$ (corresponding to a kinematic inclination of 
$38.7\pm0.5^{\circ}$).

We also use IDL routine FIT\_KINEMATIC\_PA\footnote{http://www-astro.
physics.ox.ac.uk/~mxc/software/} to measure the kinematic P.A. and systemic 
velocity, which can provide a double check on the values obtained 
above. In this code the symmetrization method 
\citep[see Appendix C of][]{Krajnovic2006} are performed, 
which minimize the differences between observed velocity field and 
a bi-antisymmetric velocity field to obtain the kinematic P.A. and 
systemic velocity. The obtained kinematic P.A. and systemic velocity 
are $122\pm0.5^{\circ}$ and $1367\pm4\ \mathrm{km}\ \mathrm{s}^{-1}$, 
respectively. While the systemic velocity is consistent with the value 
obtained from radial average, the kinematic P.A. has a difference of 
$\sim 8^{\circ}$, which could be due to the kinematic twist in the 
inner region of NGC 1042. We prefer to adopt the kinematic P.A. 
obtained from the radial average at $114.0\pm1^{\circ}$.

The disk parameters of NGC 1042 can also be measured from other observations. 
We compare these measurements with our kinematic results. Integrated HI and 
optical observations have provided several measurements of the systemic velocity 
of NGC 1042: $1371\pm2\ \mathrm{km}\ \mathrm{s}^{-1}$ \citep{Koribalski2004}, 
$1371\pm1\ \mathrm{km}\ \mathrm{s}^{-1}$ \citep{Springob2005} and 
$1372\pm3\ \mathrm{km}\ \mathrm{s}^{-1}$ 
(compilation of Hyperleda\footnote{http://leda.univ-lyon1.fr/}).
These measurements are in agreement with the systemic velocity we derived here 
within the error range. 

The photometric P.A. of NGC 1042 measured from SDSS r-band and 2MASS K-band 
images are $84^{\circ}$ \citep{Adelman-McCarthy2008} and 
$145^{\circ}$ \citep{Jarrett2003}, which are in significant 
disagreement with the kinematic P.A. obtained above. 
The discrepancy between photometric and kinematic P.A. in many other galaxies 
has been observed in GHASP, \citep{Epinat2008}, SAURON \citep{Krajnovic2008}, 
Atlas 3D \citep{Krajnovic2011} and CALIFA survey \citep{Garcia-Lorenzo2015}. 
The photometric P.A. has large systematic uncertainties associated with its 
covariance with the inclination, and with possible photometric deviations 
from axisymetry in disks which can also be color dependent. Thus we prefer 
to adapt the kinematically derived position angle.

In terms of inclination, we use the photometric axis ratios to derive the 
photometric inclination of NGC 1042. By adopting the method of \citep{Tully1988}, 
we obtain $\mathrm{i} = 40^{\circ}$ for RC3 B-band measurement \citep{deVaucouleurs1991} 
and $\mathrm{i} = 37^{\circ}$ for 2MASS K-band photometry \citep{Jarrett2003}, 
respectively. All these inclination measurements are consistent with the 
obtained kinematic inclination.

The final disk parameters used in the kinematic analysis are: center position at 
$02^{h}40^{m}24^{s}.967$, $-08^{\circ}26\arcmin00\arcsec.76$; systemic velocity at 
$1366.3\pm3\ \mathrm{km}\ \mathrm{s}^{-1}$; P.A. at $114.0\pm1^{\circ}$; and 
inclination at $38.7\pm0.5^{\circ}$.

\clearpage

\begin{deluxetable}{cc}
\tabletypesize{\scriptsize}
\tablecaption{The general parameters of NGC 1042\label{tbl-1}}
\tablewidth{0pt}
\tablehead{
\colhead{Parameters (Unit)} & 
\colhead{Value}
}
\startdata

$\alpha$(J2000.0){a}			& 02:40:24.0\\

$\delta$(J2000.0){a}			& -08:26:02\\

Type\tablenotemark{a}			& SAB(rs)cd\\

$i$\tablenotemark{a}(deg)		& 40\\

$\theta$\tablenotemark{a}(deg)		& 6\tablenotemark{e}\\

$d_{25}$\tablenotemark{a}(arcmin)	& $4.7  \times 3.6  $\\

$D_{b}$(Mpc) 				& $4.2 \pm0.7 $\\

pc/''		 			& 20\\

$M_K$\tablenotemark{c}	 		& $-19.27  \pm 0.36 $\\

$\mu_B$\tablenotemark{a}		& 23.27\\

N$_{P}$\tablenotemark{d}  		& 2\\

\enddata

\tablenotetext{a}{Coordinates, inclination ($i$), position angle ($\theta$), isophotal diameter ($d_{25}$), and effective $B$-band surface brightness ($\mu_{B}$) taken from RC3 (de Vaucouleurs et al. 1991) except when indicated.}
\tablenotetext{b}{Method to determine distance: HI 21cm Tully-Fisher (Tully et al. 2009).}
\tablenotetext{c}{From Jarrett et al. (2000).}
\tablenotetext{d}{The number of Mitchell spectrograph pointings covering the galaxy.}
\tablenotetext{e}{From Paturel et al. (2000).}

\end{deluxetable}


\begin{deluxetable}{cccccccc}
\tabletypesize{\scriptsize}
\tablecaption{Summary of VENGA Observations of NGC 1042\label{tbl-2}}
\tablewidth{0pt}
\tablehead{
\colhead{Pointing} & 
\colhead{Equatorial Coord.} & 
\colhead{Setup} &
\colhead{Dither} &
\colhead{Exposure Time} &
\colhead{N} &
\colhead{$\langle$Seeing$\rangle$} &
\colhead{$\langle$Transparency$\rangle$} 
}
\startdata
 &
\ $\alpha$\ \ \ \ \ \ \ \ \ \ \ \ \ $\delta$ &
 &
 & 
hours &
  &
$''$ & 
\\
\tableline
    & & & & & & &\\
    &                                          & red  & D1  & 2.00  & 4 & 2.20 & 0.87 \\
P1  & 2:40:26.28 -8:26:07.70  & red & D2  & 3.5  & 7 & 2.29 & 0.87 \\
      &                                        & red & D3  & 4  & 8 & 2.25 & 0.89 \\
      &                                        & blue & D1  & 0.83  & 2 & 2.00 & 0.71 \\
      &                                        & blue & D2  & 2.08  & 5 & 2.00 & 0.64 \\
      &                                        & blue & D3  & 1.67  & 4 & 2.00 & 0.73 \\
    & & & & & & &\\
\tableline
    & & & & & & &\\
    &                                         & red & D1  & 2.06  & 6 & 2.20 & 0.65 \\
P2  & 2:40:21.34 -8:25:56.10 & red & D2  & 2.50  & 5 & 2.52 & 0.67 \\
    &                                         & red & D3  & 3.50  & 7 & 1.90 & 0.68 \\
     &                                        & blue & D1  & 2.08  & 5 & 2.58 & 0.68 \\
     &                                        & blue & D2  & 1.25  & 3 & 2.93 & 0.69 \\
     &                                        & blue & D3  & 1.25  & 3 & 1.63 & 0.65 \\
    & & & & & & &\\
\enddata
\end{deluxetable}

\clearpage

\begin{deluxetable}{ccccc}
\tabletypesize{\scriptsize}
\tablecaption{Fitted Emission Lines in NGC 1042\label{tbl-3}}
\tablewidth{0pt}
\tablehead{
\colhead{Transition} & 
\colhead{Wavelength} & 
\colhead{Median S/N} &
\colhead{Fraction - $5\sigma$} &
\colhead{Fraction - $3\sigma$}
}
\startdata
 &
\AA\ &
      &
    & 
\\
\tableline
 &  &  &     & \\

[OII]\tablenotemark{a}               & 3726.03 & 9.7  & 0.84     & 0.94\\

[OII]\tablenotemark{a}                  & 3728.73 & -  & -     & -\\

[NeIII]               & 3868.69 & 1.4  & 0.02     & 0.07\\

[NeIII]               & 3967.40 & 0.5  & $<$0.01     & 0.01\\

H8                    & 3889.06 & 1.7 & 0.09     & 0.20\\

H$\epsilon$     & 3970.08 & 1.7  & 0.08     & 0.18\\

H$\delta$        & 4101.73 & 2.3  & 0.18     & 0.34\\

H$\gamma$    & 4340.47 & 5.3  & 0.52     & 0.77\\

[OIII]                & 4363.15 & 0.9  & $<$0.01     & $<$0.01\\

HeII                 & 4685.74 & 0.6  & $<$0.01     & 0.01\\

H$\beta$        & 4861.32 & 16 & 0.91  & 0.95\\

[OIII]                & 4958.83 & 2.8  & 0.23   & 0.42\\

[OIII]               & 5006.77 & 8.6  & 0.78  & 0.90\\

[NI]\tablenotemark{a}                  & 5197.90 & 1.3  & $<$0.01     & 0.06\\
           
[NI]\tablenotemark{a}                  & 5200.39 & -  & -    & - \\

[NII]                & 6547.96 & 6.9  & 0.63  & 0.80\\

H$\alpha$      & 6562.80 & 38.9 & 0.98  & 0.99\\

[NII]                & 6583.34 & 21.2 & 0.94  & 0.97\\

[SII]                 & 6716.31 & 13.3 & 0.93  & 0.96\\

[SII]                & 6730.68 & 9.2  & 0.83  & 0.94\\

\enddata
\tablenotetext{a}{Since we cannot resolve the [OII]$\lambda$3727 and
  [NI]$\lambda$5200 doublets, we report the median S/N and fraction of
  the observed area in which the lines are significantly detected for
  the sum of the two doublet components.}

\end{deluxetable}

\end{document}